\documentclass[12pt,journal,onecolumn,draftcls]{IEEEtran}

\pdfoutput = 1

\usepackage[cmex10]{amsmath}
\usepackage{dsfont}
\usepackage{algorithm}
\usepackage{algorithmic}
\usepackage[dvips]{graphicx}

\DeclareMathOperator*{\argmax}{arg\,max}
\DeclareMathOperator*{\argmin}{arg\,min}
\DeclareMathOperator{\diag}{diag}

\usepackage{color,ulem}

\usepackage{lineno}

\begin{document}

\title{
An algorithm for constrained
one-step inversion of spectral CT data
}

\author{Rina Foygel Barber,
\thanks{Rina Foygel Barber is with the Department of Statistics,
The University of Chicago, 5734 S. University Ave., Chicago, IL 60637, USA.
rina@uchicago.edu}
Emil Y. Sidky,
\thanks{Emil Y. Sidky and Xiaochuan Pan are with the Department of Radiology,
The University of Chicago, 5841 S. Maryland Ave., Chicago, IL 60637, USA.
e-mail: \{sidky,xpan\}@uchicago.edu}
Taly Gilat Schmidt,
\thanks{Taly Gilat Schmidt is with the Department of Biomedical Engineering,
Marquette University, Milwaukee, WI 53233, USA.
taly.gilat-schmidt@marquette.edu}
and Xiaochuan Pan}

\maketitle

\begin{abstract}
We develop a primal-dual algorithm that allows for one-step inversion of spectral CT
transmission photon counts data to a basis map decomposition. The algorithm allows
for image constraints to be enforced on the basis maps during the inversion. The derivation
of the algorithm makes use of a local upper bounding quadratic approximation to generate
descent steps for non-convex spectral CT data discrepancy terms, combined with
a new convex-concave optimization algorithm. Convergence of
the algorithm is demonstrated on simulated spectral CT data.  Simulations with noise
and anthropomorphic phantoms show examples of how to employ the constrained one-step
algorithm for spectral CT data.
\end{abstract}

\section{Introduction}

The recent research activity in photon-counting detectors has motivated a
resurgence in the investigation of spectral computed tomography (CT). Photon-counting
detectors detect individual X-ray quanta and the electronic pulse signal generated
by these quanta has a peak amplitude proportional to the photon energy \cite{Taguchi2013}.
Thresholding these amplitudes allows for coarse energy resolution of the X-ray photons,
and the transmitted flux
of X-ray photons can be measured simultaneously in a number of energy windows.
Theoretically, the energy-windowed transmission measurements can be exploited to reconstruct
quantitatively the X-ray attenuation map of the subject being scanned \cite{Alvarez1976}.
The potential benefits are reduction of beam-hardening artifacts, and improved
contrast-to-noise ratio (CNR), signal-to-noise ratio (SNR),
and quantitative imaging \cite{Taguchi2013,Shikhaliev2008b,Schmidt2009,Alessio2013}.
For photon-counting detectors where the number of energy windows  can be three or
greater, the new advantage with respect to quantitative imaging is the ability to image
contrast agents that possess a K-edge in the
diagnostic X-ray energy range \cite{Roessl2007,Schlomka2008,Roessl2011a,Cormode2010,Roessl2011b,Schirra2013}.

Use of energy information in X-ray CT has been proposed almost
since the conception of CT itself \cite{Hounsfield1973}.
Dual-energy
CT acquires transmission intensity at either two energy windows or for two different
X-ray source spectra. Despite the extremely
coarse energy-resolution, the technique is effective because for many materials
only two physical processes, photo-electric effect and Compton scattering,
dominate X-ray attenuation in the diagnostic energy range \cite{Alvarez1976}. Within the context of
dual-energy, the processing methods of energy-windowed intensity data have been
classified in two broad categories: pre-reconstruction and post-reconstruction \cite{Maass2009}.
The majority of processing methods for multi-window data also fall into these categories.

In pre-reconstruction processing of the multi-energy data, the X-ray attenuation map
is expressed as a sum of terms based on physical process or basis materials \cite{Alvarez1976}.
The multi-energy data are converted to sinograms of the basis maps, then any image
reconstruction technique can be employed to invert these sinograms. The basis maps can
subsequently be combined to obtain images of other desired quantities: estimated
X-ray attenuation maps at a single energy, material maps, atomic number, or electron density maps.
The main advantage of pre-reconstruction
processing is that beam-hardening artifacts can be avoided, because consistent sinograms
of the basis maps are estimated prior to image reconstruction.  Two major challenges for pre-reconstruction
methods are the need to calibrate the spectral transmission model and
to acquire registered projections. Photon-counting detectors 
ease the implementation of projection registration, because multiple energy-thresholding
circuits can operate on the same detection element signal. Accounting for detection physics
and spectral calibration by data pre-processing or incorporation directly in the image
reconstruction algorithm remains a challenge for photon-counting detectors \cite{Taguchi2013}.

For post-reconstruction processing, the energy-windowed transmission data are processed
by the standard negative logarithm to obtain approximate sinograms of a weighted energy-averaged
attenuation map followed by standard image reconstruction.  The resulting images can be
combined to obtain approximate estimates of images of the same physical quantities
as mentioned for the pre-reconstruction processing \cite{Brooks1977}. The advantage of post-reconstruction
processing is that it is relatively simple, because it is only a small modification
on how standard CT data are processed and there is no requirement of projection registration.
The down-side, however, is that the images corresponding to each energy-window are
susceptible to beam-hardening artifacts because the negative logarithm processed data
will, in general, not be consistent with the projection of any object.

A third option for the processing of spectral CT data, however, does exist, which due 
to difficulties arising from the nonlinearity of the attenuation of polychromatic X-rays when passing 
through an object,
is much less common than either pre- or post-reconstruction methods: direct estimation
of basis maps from energy-windowed transmission data.
This approach, labeled the
one-step approach in the remainder of the article,
has the advantages that the spectral transmission model is treated exactly, there is no
need for registered projections, and constraints on the basis maps can be incorporated
together with the fitting of the spectral CT data.
The main difficulty of the one-step approach is that it necessitates an iterative algorithm
because the corresponding transmission data model is too complex for analytic solution,
at present. Iterative image reconstruction (IIR) has been applied to spectral CT
in order to address the added complexity of the data
model \cite{Fessler2002,Elbakri2002,Chung2010,Cai2013,Zhang2014,Long2014,Sawatsky2014,Nakada2015}.

In this work, we develop a framework that addresses one-step image reconstruction in spectral CT
allowing for non-smooth convex constraints to be applied to the basis maps. We demonstrate
the algorithm with the use of total variation (TV) constraints, but the framework allows
for other constraints such as non-negativity, upper bounds, and sum bounds applied
to either the basis maps or to a composite image such as an estimated mono-chromatic attenuation
map.

We draw upon recent developments in large-scale first-order algorithms
and adapt them to incorporate the non-linear model for spectral CT
to optimize the data-fidelity of the estimated image by minimizing
the discrepancy between the observed and estimated data.  We present an algorithm
framework for constrained optimization, deriving algorithms for minimizing the
data discrepancy
based on least-squares fitting and on a transmission Poisson likelihood model. As previously
mentioned, the framework
admits many convex constraints that can be exploited to stabilize image reconstruction from
spectral CT data. Section \ref{sec:onestep} presents the constrained optimization for one-step
spectral CT image reconstruction; Sec. \ref{sec:algorithm} presents a convex-concave
primal-dual algorithm that addresses the non-convex data discrepancy term arising from the
non-linear spectral CT data model; and Sec. \ref{sec:results} demonstrates the proposed
algorithm with simulated spectral CT transmission data.

\section{One-step image reconstruction for spectral CT}
\label{sec:onestep}

\subsection{Spectral CT data model}

For the present work, we employ a basic spectral model for the 
energy-windowed transmitted X-ray intensity along a ray $\ell$, where the 
transmitted X-ray intensity in the energy window $w$ for ray $\ell$ is given by
\begin{linenomath}
\begin{equation*}
I_{w\ell} = \int_{E}  S_{w \ell}(E) \exp \left[ - \int_{t\in\ell}
\mu(E,\vec{r}(t)) \, \mathsf{d}t \right]\,\mathsf{d}E.
\end{equation*}
\end{linenomath}
Here $\int_{t\in\ell}$ denotes that we are integrating along the ray $\ell$
while $\int_E$ integrates over the range of energy; $S_{w\ell}(E)$ is 
the product of the X-ray beam spectrum intensity and detector sensitivity for
the energy window $w$ and transmission ray $\ell$ at energy $E$; and
$\mu(E,\vec{r})$ is the linear X-ray attenuation coefficient for energy $E$ at the spatial
location $\vec{r}$. Let $I^{(0)}_{w \ell} $ be the transmitted intensity
in the setting where no object is present between the X-ray beam and the detector (i.e.~attenuation
is set to zero), given by
\begin{linenomath}
\begin{equation}
\notag
I^{(0)}_{w \ell} =\int_E S_{w \ell}(E) \, \mathsf{d}E; \;
s_{w \ell}(E) =  S_{w \ell}(E)/ I^{(0)}_{w \ell}.
\end{equation}
\end{linenomath}
Then we can write
\begin{linenomath}
\begin{equation}\label{spectralModel}
I_{w\ell} = I^{(0)}_{w \ell} \int_E s_{w \ell}(E)
\exp \left[ - \int_{t\in\ell} \mu(E,\vec{r}(t)) \, \mathsf{d}t\right] \, \mathsf{d}E,
\end{equation}
\end{linenomath}
where $s_{w \ell}(E)$ represents the normalized energy distribution of X-ray intensity and
detector sensitivity.
Image reconstruction for spectral CT aims to recovery the a complete energy-dependent
linear attenuation map $\mu(E,\vec{r})$ from intensity measurements $I_{w \ell}$ in all
windows $w$ and rays $\ell$ comprising the X-ray projection data set.

Throughout the article we use the convention that $N_x$ is the dimension of the discrete
index $x$. For example, the spectral CT data set consists of $N_w$ energy windows and
$N_\ell$ transmission rays.

This inverse problem is simplified by exploiting the fact that the energy-dependence
of the X-ray attenuation coefficient can be represented efficiently by a low-dimensional
expansion. For the present work, we employ the basis material expansion
\begin{linenomath}
\begin{equation}
\label{materials}
\mu(E, \vec{r}) =\sum_m \mu_m(E) f_m(\vec{r})= \sum_m \left( \frac{\mu_m(E)}{\rho_m} \right) \rho_m f_m(\vec{r}),
\end{equation}
\end{linenomath}
where $\rho_m$ is the density of material $m$;
the X-ray mass attenuation coefficient
$\mu_m(E)/\rho_m$ are available from the 
national institute of standards and technology (NIST) report by Hubbell and Seltzer \cite{Hubbell1995};
and $f_m(\vec{r})$ is the fractional density map of material $m$ at location $\vec{r}$.
For the present spectral CT image reconstruction problem, we aim to recover
$f_m(\vec{r})$, which we refer to as the material maps.

Proceeding with the spectral CT model, we discretize
the material maps $f_m(\vec{r})$ by use of an expansion set
\begin{linenomath}
\begin{equation}
\notag
f_m(\vec{r}) = \sum_k^{N_k} f_{k m} \phi_k^\text{(map)}(\vec{r}) ,
\end{equation}
\end{linenomath}
where $\phi_k^\text{(map)}(\vec{r})$
are the representation functions for the  material maps,
respectively.  For the 2D/3D image representation standard pixels/voxels are employed, that is,
$k$ indexes the pixels/voxels.
With the spatial expansion set, the line integration over the material maps is
represented by a matrix $X$ with entry $X_{\ell k}$ measuring the length of the intersection
between ray $\ell$ and pixel $k$:
\begin{linenomath}
\begin{equation}
\notag
\int_{t\in\ell} \mu(E,\vec{r}(t))\,\mathsf{d}t =  \sum_{m k} \mu_m(E) X_{\ell k} f_{k m},
\end{equation}
\end{linenomath}
where formally we can calculate
\begin{linenomath}
\begin{equation}
\notag
X_{\ell k} = \int_{t\in\ell} \phi_k^\text{(map)} (\vec{r} ) \, \mathsf{d}t.
\end{equation}
\end{linenomath}
This integration
results in the standard line-intersection method for the pixel/voxel basis.

The discretization of the integration over energy $E$ in Eq. (\ref{spectralModel})
is perform by use of a Riemann sum approximation.
\begin{linenomath}
\begin{multline}
\notag
I_{w\ell} 
= I^{(0)}_{w \ell} \int_{E} s_{w \ell}(E)
\exp \left[ - \int_{t\in\ell} \mu(E,\vec{r}) \, \mathsf{d}t \right] \, \mathsf{d}E\\
\approx I^{(0)}_{w \ell} \sum_i \Delta E_i s_{w \ell}(E_i)
\exp \left[ - \int_{t\in \ell} \mu(E_i,\vec{r})\, \mathsf{d}t \right]
= I^{(0)}_{w \ell} \sum_i s_{w \ell i}
\exp \left[ -\sum_{m k} \mu_{m i} X_{\ell k} f_{k m} \right],
\end{multline}
\end{linenomath}
where $i$ indexes the discretized energy $E$ and
\[s_{w \ell i} = \Delta E_i s_{w \ell}(E_i)\text{ and }\mu_{m i} = \mu_m(E_i).\]
With the Riemann sum approximation we normalize the discrete window
spectra,
\begin{linenomath}
\begin{equation}
\notag
\sum_i s_{w \ell i} = 1.
\end{equation}
\end{linenomath}
Modeling photon-counting detection, we express X-ray
incident and transmitted spectral fluence in terms of numbers of photons
per ray $\ell$ (as before, the ray $\ell$ identifies the source detector-bin combinations) and energy window
$w$:
\begin{linenomath}
\begin{equation}
\label{meanModel}
\hat{c}_{w \ell} = N_{w \ell} \sum_i s_{w \ell i}
\exp \left[ -\sum_{m k} \mu_{m i} X_{\ell k} f_{k m} \right],
\end{equation}
\end{linenomath}
where $N_{w \ell}$ is the incident spectral fluence
and $\hat{c}_{w \ell}$ is interpreted as a mean transmitted fluence.
Note that in general the right hand side of Eq. (\ref{meanModel}) evaluates
to a non-integer value and as a result the left hand side variable cannot
be assigned to an integer as would be implied by reporting transmitted fluence
in terms of numbers of photons. This inconsistency is rectified by
interpreting the left hand side variable, $\hat{c}_{w \ell}$, as an expected value.

\subsection{Constrained optimization for one-step basis decomposition}
\label{sec:constrainedOpt}

For the purpose of developing one-step image reconstruction of the basis
material maps from transmission counts data, we formulate a constrained
optimization involving minimization of a non-convex data-discrepancy objective
function subject to convex constraints. The optimization problem of interest
takes the following form
\begin{linenomath}
\begin{equation}
\label{grandOpt}
f^* = \argmin_{f} \left\{
\sum_{w \ell} D(c_{ w \ell}, \hat{c}_{w \ell} (f) ) + \sum_i \delta(P_i)
\right\},
\end{equation}
\end{linenomath}
where $c_{ w \ell}$ are the measured counts in energy window $w$ and ray $\ell$;
$D(\cdot,\cdot)$ is a generic data discrepancy objective function; and the
indicator functions $\delta(P_i)$ enforce the convex constraints
$f \in P_i$, the $P_i$ are convex sets corresponding to the desired constraints (for instance,
nonnegativity of the material maps). The indicator function is defined
\begin{linenomath}
\begin{equation}
\label{indicatordef}
\delta(P) = \begin{cases}
0 & f \in P\\
\infty & f \notin P
\end{cases}.
\end{equation}
\end{linenomath}
Use of constrained optimization with TV constraints is demonstrated in
Sec.~\ref{sec:results}.

\paragraph*{Data discrepancy functions}

For the present work, we consider two data discrepancy functions: transmission Poisson
likelihood (TPL) and least-squares (LSQ)
\begin{linenomath}
\begin{align}
\label{data-TPL}
D_\text{TPL}(c, \hat{c} (f) ) &=
\sum_{w \ell} \left[ \hat{c}_{w \ell} (f) - c_{w \ell} -
c_{w \ell} \log \left( \hat{c}_{w \ell} (f)/ c_{w \ell} \right) \right] \\
\label{data-LSQ}
D_\text{LSQ}(c, \hat{c} (f) ) &=
\frac{1}{2} \sum_{w \ell} \left[ \log ( c_{w \ell} ) -\log \left( \hat{c}_{w \ell} (f) \right) \right]^2.
\end{align}
\end{linenomath}
The TPL data discrepancy function is derived from the negative log likelihood of
a stochastic model of the counts data
\begin{linenomath}
\begin{equation}
\notag
c_{w \ell} \sim\mathsf{Poisson}\left( \hat{c}_{w \ell} (f) \right),
\end{equation}
\end{linenomath}
that is, minimizing $D_\text{TPL}$ is equivalent to maximizing the likelihood.
Note that in defining $D_\text{TPL}$ we have subtracted a term independent of $f$ from
the negative log likelihood so that $D_\text{TPL}$ is zero when $c =\hat{c}(f)$, and positive otherwise.
From a physics perspective, the important difference between these two data discrepancy functions is
how they each weight the individual measurements; the LSQ function treats all measurements
equally while the TPL function gives greater weight to higher count measurements.
We point out this property to emphasize that the TPL data discrepancy
can be useful even when there are data inconsistencies due to other
physical factors besides the stochastic nature of the counts measurement.
This alternate weighting is also achieved without introducing additional
parameters as would be the case for a weighted quadratic data discrepancy.
From a mathematics perspective, both data functions are convex functions
of $\hat{c}_{w \ell} (f)$, but they are non-convex functions of $f$. It is
the non-convexity with respect to $f$ that drives the main theoretical 
and algorithmic development of this work.
Although we consider only these two data fidelities, the same methods can be applied
to other functions.

\paragraph*{Convex constraints}

The present algorithm framework allows for convex constraints that may improve reconstruction
of the basis material maps. In Eq. (\ref{grandOpt}) the constraints are coded with indicator
functions, but here we express the constraints by the inequalities that define the convex
set to which the material maps are restricted.
When the basis materials are identical to the materials actually present in the subject,
the basis maps can be highly constrained.
Physically, the fractional densities represented by each material map must take on a value
between zero and one, and the corresponding constraint is
\begin{linenomath}
\begin{equation}
\label{bounds}
0 \le f_{m k} \le 1.
\end{equation}
\end{linenomath}
Similarly, the sum of the fractional densities cannot be greater than one, leading to
a constraint on the sum of material maps
\begin{linenomath}
\begin{equation}
\label{sum}
\sum_m f_{m k} \le 1.
\end{equation}
\end{linenomath}
Care must be taken, however, in using these bound and sum constraints when the basis materials
used for computation are not the same as the materials actually present in the scanned object.
The bounds on the material maps
and their sum must likely be loosened, and therefore they may not be as effective.

In medical imaging, where multiple soft tissues comprise the subject, it is standard
to employ a spectral CT materials basis which does not include many of the tissue/density
combinations present.
The reason for this is that soft tissues such as muscle, fat, brain, blood, etc., all have
attenuation curves similar to water, and recovering each of these soft tissues individually becomes an
extremely ill-posed inverse problem. For spectral CT, it is common to employ a two-material
expansion set, such as bone and water, and possibly a third material for representing
contrast agent that has a K-edge in the diagnostic X-ray energy range. The displayed
image can then be the basis material maps or
the estimated X-ray attenuation map for a single energy $E$, also known
as a monochromatic image
\begin{linenomath}
\begin{equation}
\label{monoimage}
f^\text{(mono)}_k(E) = \sum_m \left( \frac{\mu_m(E)}{\rho_m} \right)
\rho_m f_{m k}.
\end{equation}
\end{linenomath}
A non-negativity constraint can be applied to the monochromatic image
\begin{linenomath}
\begin{equation}
\notag
f^\text{(mono)}_k(E) \ge 0
\end{equation}
\end{linenomath}
at one or more energies. This constraint makes physical sense even when the basis
materials are not the same as the materials in the subject.

Finally, we formulate $\ell_1$-norm constraints on the gradient magnitude images,
also known as the total variation,
in order to encourage gradient magnitude sparsity in either the basis material maps
or the monochromatic image.
In applying TV constraints to the basis material maps,
we allow for different constraint values $\gamma_m$ for each material
\begin{linenomath}
\begin{equation}
\notag
\|f_m\|_\text{TV} \equiv \left\| \left( | \nabla f_{m } | \right) \right\|_1 \le \gamma_m,
\end{equation}
\end{linenomath}
where $\nabla$ represents the finite-differencing approximation to the gradient,
and we use $| \cdot |$ to represent a spatial magnitude operator
so that $|  \nabla f_{m } |$ is the gradient magnitude image (GMI) of material map $m$.
Similarly, a TV constraint can be formulated so that it applies to the monochromatic
image at energy $E$
\begin{linenomath}
\begin{equation}
\notag
\|f^\text{(mono)}(E) \|_\text{TV} \equiv
\left\| \left( | \nabla f^\text{(mono)}(E) | \right) \right\|_1
\le \gamma_\text{mono}(E),
\end{equation}
\end{linenomath}
where the constraint can be applied at one or more values of $E$.

The constraints involving TV of the material maps and the monochromatic
image are specifically in Sec.~\ref{sec:results}.
Many other convex constraints can be incorporated into the presented framework
such as constraints on a generalized TV computed from multiple monochromatic images
\cite{Rigie2015}.

\section{A first-order algorithm for spectral CT constrained optimization}
\label{sec:algorithm}

The proposed algorithm derives from the primal-dual algorithm of Chambolle
and Pock (CP) \cite{Chambolle2011,Pock2011,Sidky2012}.
Considering the general constrained optimization form in Eq. (\ref{grandOpt}),
the second term coding the convex constraints can be treated in the same way
as shown in Refs. \cite{Sidky2014,Jorgensen2015}.
The main algorithmic development, presented here, is the generalization and
adaptation of CP's primal-dual algorithm to the minimization of the data discrepancy
term, the first term of Eq. (\ref{grandOpt}).  We derive
the data fidelity steps specifically focusing on the deriving steps for $D_\text{TPL}$
and $D_\text{LSQ}$.

\paragraph*{Optimizing the spectral CT data fidelity}
We first sketch the main developments of the algorithm for minimizing the non-convex
data discrepancy terms, and then explain each step in detail. The overall design
of the algorithm is comprised of two nested iteration loops. The outer iteration loop
involves derivation of a convex quadratic upper bound to the local quadratic
Taylor expansion about the current estimate for the material maps. The inner iteration
loop takes descent steps for the current quadratic upper bound.
Although the algorithm construction formally involves two nested
iteration loops, in practice the number of inner loop iterations is set to one.
Thus, effectively the algorithm consists only of a single iteration loop where
a re-expansion of the data discrepancy term is performed at every iteration.

The local convex quadratic upper bound, used to generate descent steps for
the non-convex data discrepancy terms, does not fit directly with the generic primal-dual
optimization form used by CP. A convex-concave generalization to the CP primal-dual
algorithm is needed. The resulting algorithm
called \underline{m}irr\underline{o}red \underline{c}onvex-\underline{c}onc\underline{a}ve
(MOCCA) algorithm is presented in detail in Ref. \cite{Barber2015}. For the one-step
spectral CT image reconstruction algorithm we present:
the local convex quadratic upper bound, a short description of MOCCA and its
application in the present context, preconditioning, and convergence checks for the spectral CT image
reconstruction algorithm.

\subsection{A local convex quadratic upper bound to the spectral CT data discrepancy terms}

\subsubsection{Quadratic expansion}
We carry out the deriviations on $D_\text{LSQ}$ and $D_\text{TPL}$ in parallel.
The local quadratic expansion for each of these data discrepancy terms about
the material maps $f=f_0$ is
\begin{linenomath}
\begin{align}
\label{quadexp}
&L(f)  \approx L\left(f_0\right) + 
\left(f-f_0\right)^\top  \nabla_f L\left(f_0\right) +
\frac{1}{2} \left(f-f_0\right)^\top  \nabla^2_f L\left(f_0\right) \left(f-f_0\right), \\
\text{where} & \notag \\
&L(f) = D(c , \hat{c} (f)).  \notag
\end{align}
\end{linenomath}
To obtain the desired expansions, we need expressions for the gradient and Hessian 
of each data discrepancy. The gradient of $L_\text{TPL}(f)$ is derived
explicitly in Appendix \ref{sec:derivatives}; we do not show the details for the other
derivations.
The data discrepancy gradients are:
\begin{linenomath}
\begin{align}
\label{TPLgrad}
\nabla_f L_\text{TPL}(f) &= Z^\top A(f)^\top r(f), \\
\label{LSQgrad}
\nabla_f L_\text{LSQ}(f) &= Z^\top A(f)^\top r^\text{(log)}(f),
\end{align}
\end{linenomath}
where $r$ and $r^\text{(log)}$ denote the residuals in terms of counts or log counts:
\begin{linenomath}
\begin{align}
\label{resid}
r_{w \ell}(f) &= c_{w \ell} - \hat{c}_{w \ell}(f), \\
\label{logresid}
r^\text{(log)}_{w \ell}(f) &= \log(c_{w \ell}) - \log \left( \hat{c}_{w \ell}(f)\right);
\end{align}
\end{linenomath}
$Z$ represents the combined linear transform that accepts material maps,
performs projection, and then combines the resulting sinograms to form
monochromatic sinograms at energy $E_i$:
\begin{linenomath}
\begin{equation}
\label{zdef}
Z_{\ell i, m k} = \mu_{m i} X_{\ell k};
\end{equation}
\end{linenomath}
 and $A(f)$ is a term that results from the gradient of the logarithm
of the estimated counts $\log \hat{c}(f)$:
\begin{linenomath}
\begin{align}
\label{adef}
A_{w \ell, \ell^\prime i}(f) = & \frac{s_{w \ell i} \exp \left[- (Zf)_{\ell i} \right]}
{\sum_{i^\prime} s_{w \ell i^\prime} \exp \left[- (Zf)_{\ell i^\prime} \right]}
 \mathbf{I}_{\ell \ell^\prime} \\
\notag
 \mathbf{I}_{\ell \ell^\prime} = &
\begin{cases}
0 & \ell \neq \ell^\prime \\
1 & \ell = \ell^\prime
\end{cases}
.
\end{align}
\end{linenomath}
Using the same variable and linear transform definitions, the expressions for the two Hessians 
are
\begin{linenomath}
\begin{align}
\label{TPLlap}
\nabla^2_f L_\text{TPL}(f) &= -Z^\top \diag \left(A(f)^\top r(f)\right) Z +
Z^\top A(f)^\top \diag \left( \hat{c}(f) + r(f) \right) A(f) Z, \\
\label{LSQlap}
\nabla^2_f L_\text{LSQ}(f) &=  -Z^\top \diag \left(A(f)^\top r^\text{(log)}(f)\right) Z +
Z^\top A(f)^\top \diag \left( 1 + r^\text{(log)}(f) \right) A(f) Z. 
\end{align}
\end{linenomath}
Substituting either Eq. (\ref{TPLlap}) or (\ref{LSQlap}) for the Hessian
and either  Eq. (\ref{TPLgrad}) or (\ref{LSQgrad}) for the gradient into
the Taylor expansion in Eq. (\ref{quadexp}), yields the quadratic approximation
to the data discrepancy terms of interest. This quadratic is in general non-convex
because both Hessian expressions can have negative values.

\subsubsection{A local convex upper bound to $L(f)$}
The key to deriving a local convex upper bound to the quadratic expansion
of $L(f)$ is to split the Hessian expressions into positive and negative
components.  Setting the negative components to zero and substituting this
thresholded Hessian into the Taylor's expansion, yields a quadratic
term with non-negative curvature. (As an aside, a tighter convex local quadratic
upper bound would be attained by diagonalizing the Hessian and forming
a positive semi-definite Hessian by keeping eigenvectors corresponding to
only non-negative eigenvalues in the eigenvalue decomposition, but for
realistic sized tomography configurations such an eigenvalue decomposition is
impractical.)
The algebraic steps for splitting the Hessian into positive and negative components
in the form
\begin{linenomath}
\begin{equation}
\notag
\nabla^2_f L(f) = \nabla^2_+L(f) - \nabla^2_-L(f),
\end{equation}
\end{linenomath}
where $\nabla^2_+L(f)$ and
$\nabla^2_-L(f)$  are both positive semidefinite (see Appendix \ref{sec:posdef} for
more details).
The resulting split expressions are:
\begin{linenomath}
\begin{align}
\label{TPLposH}
\nabla^2_+ L_\text{TPL}(f) &= Z^\top \diag \left(A(f)^\top r_-(f)\right) Z +
Z^\top A(f)^\top \diag \left( \hat{c}(f) - r_-(f) \right) A(f) Z, \\
\label{TPLnegH}
\nabla^2_- L_\text{TPL}(f) &= Z^\top \diag \left(A(f)^\top r_+(f)\right) Z -
Z^\top A(f)^\top \diag \left( r_+(f) \right) A(f) Z,
\end{align}
\end{linenomath}
and
\begin{linenomath}
\begin{align}
\label{LSQposH}
\nabla^2_+ L_\text{LSQ}(f) &=  Z^\top \diag \left(A(f)^\top r_-^\text{(log)}(f)\right) Z +
Z^\top A(f)^\top \diag \left( 1 - r_-^\text{(log)}(f) \right) A(f) Z,\\
\label{LSQnegH}
\nabla^2_- L_\text{LSQ}(f) &=  Z^\top \diag \left(A(f)^\top r_+^\text{(log)}(f)\right) Z -
Z^\top A(f)^\top \diag \left(r_+^\text{(log)}(f) \right) A(f) Z, 
\end{align}
\end{linenomath}
where
\begin{linenomath}
\begin{equation}
\notag
r(f) = r_+(f) - r_-(f), \; \; r_+(f) = \max \left[ r(f), 0 \right]
\text{ and } r_-(f) = \max \left[ -r(f), 0 \right],
\end{equation}
\end{linenomath}
and similarly
\begin{linenomath}
\begin{equation}
\notag
r^\text{(log)}(f) = r^\text{(log)}_+(f) - r^\text{(log)}_-(f), \; \;
r_+^\text{(log)}(f) = \max \left[ r^\text{(log)}(f), 0 \right]
\text{ and } r^\text{(log)}_-(f) = \max \left[ -r^\text{(log)}(f), 0 \right].
\end{equation}
\end{linenomath}

To summarize, the expression for the convex local upper bound to the
quadratic approximation in Eq. (\ref{quadexp}) is
\begin{linenomath}
\begin{equation}
\label{qup}
Q\left(c,f_0; f\right)  =  L\left(f_0\right) + 
\left(f-f_0\right)^\top  \nabla_f L\left(f_0\right) +
\frac{1}{2} \left(f-f_0\right)^\top  \nabla^2_+ L\left(f_0\right) \left(f-f_0\right),
\end{equation}
\end{linenomath}
where $\nabla^2_+$ is used instead of $\nabla^2_f$ in the quadratic term.
Here $Q$
depends parametrically on the counts data $c$, through the function $L$
(see Eq. (\ref{quadexp})), and the expansion center $f_0$.
The gradients of $L$ at $f_0$ are obtained from Eqs. (\ref{TPLgrad}) and (\ref{LSQgrad}),
and the Hessian upper bounds are available from Eqs. (\ref{TPLposH}) and (\ref{LSQposH}).
Note that the quadratic expression in Eq. (\ref{qup}) is not necessarily an upper bound of the data
discrepancy functions, even locally, because we bound only the quadratic expansion.
We employ
the convex function $Q\left(c,f_0; f\right)$ combined with convex constraints
to generate descent steps for the generic non-convex optimization problem specified
in Eq. (\ref{grandOpt}).

\subsection{The motivation and application of MOCCA}

\subsubsection{Summary of the Chambolle-Pock (CP) primal-dual framework}
The generic convex optimization addressed in Ref. \cite{Chambolle2011} is
\begin{linenomath}
\begin{equation}
\label{cpgen}
x^\star = \argmin_x \left\{ F(K x) + G(x) \right\},
\end{equation}
\end{linenomath}
where $F$ and $G$ are convex, possibly non-smooth, functions and $K$ is
a matrix multiplying the vector $x$. The ability to handle non-smooth
convex functions is key for addressing the convex constraints of
Eq. (\ref{grandOpt}).  In the primal-dual picture this minimization
is embedded in a larger saddle point problem
\begin{linenomath}
\begin{equation}
\label{saddlept}
\min_x \max_y \left\{ y^\top K x - F^*(y) + G(x) \right\},
\end{equation}
\end{linenomath}
using the Legendre transform or convex conjugation
\begin{linenomath}
\begin{equation}
\label{legendre}
F^*(y) = \max_{x} \left\{ x^\top y - F(x) \right\},
\end{equation}
\end{linenomath}
and the fact that
\begin{linenomath}
\begin{equation}
\label{doublestar}
F(x) = F^{**}(x) =  \max_{y} \left\{ y^\top x - F^*(y) \right\}
\end{equation}
\end{linenomath}
if $F$ is a convex function. The CP primal-dual algorithm of interest
solves Eq. (\ref{saddlept}) by iterating on the following steps
\begin{linenomath}
\begin{align}
\label{dualstep}
y^{(n+1)} & =  \argmin_{y^\prime} \left\{F^*(y^\prime)
+ \frac{1}{2 \sigma} \|y^{(n)}+\sigma K \bar{x}^{(n)} - y^\prime \|^2_2
\right\}\\
\label{primalstep}
x^{(n+1)} & =  \argmin_{x^\prime} \left\{G(x^\prime)
+ \frac{1}{2 \tau} \|x^{(n)}-\tau K^\top y^{(n+1)} - x^\prime \|^2_2 \right\}\\
\label{predictor}
\bar{x}^{(n+1)} &=  2x^{(n+1)} -x^{(n)},
\end{align}
\end{linenomath}
where $n$ is the iteration index; $\sigma>0$ and $\tau>0$ are the primal
and dual step sizes, respectively, and these step sizes must satisfy
the inequality
\begin{linenomath}
\begin{equation}
\notag
\sigma \tau < \frac{1}{\|K\|^2_2}
\end{equation}
\end{linenomath}
where $\|K\|_2$ is the largest singular value of $K$.
Because this algorithm solves the saddle point problem, Eq. (\ref{saddlept}),
one obtains the solution to the primal problem, Eq. (\ref{cpgen}) along
with its Fenchel dual
\begin{linenomath}
\begin{equation}
\label{cpgen_dual}
y^\star = \argmax_x \left\{ -F^*(y) - G^*(-K^\top y) \right\}.
\end{equation}
\end{linenomath}
The fact that both Eqs. (\ref{cpgen}) and (\ref{cpgen_dual}) are solved simultaneously
provides a convergence check: the primal-dual gap, the difference between
the objective functions of Eqs. (\ref{cpgen}) and (\ref{cpgen_dual}), tends
to zero as the iteration number increases.

In some settings, the requirement $\sigma \tau < \frac{1}{\|K\|^2_2}$ may be impractical
or too conservative, and the CP algorithm can instead be implemented with diagonal matrices
$\Sigma$ and $T$ in place of $\sigma$ and $\tau$ \cite{Pock2011}, with
the condition $\|\Sigma^{1/2}K T^{1/2}\|< 1$ and the revised steps
\begin{linenomath}
\begin{align}
\label{pcdualstep}
y^{(n+1)} & =  \argmin_{y^\prime} \left\{F^*(y^\prime)
+ \frac{1}{2 } \|y^{(n)}+\Sigma K \bar{x}^{(n)} - y^\prime \|^2_{\Sigma^{-1}}
\right\}\\
\label{pcprimalstep}
x^{(n+1)} & =  \argmin_{x^\prime} \left\{G(x^\prime)
+ \frac{1}{2 } \|x^{(n)}- T K^\top y^{(n+1)} - x^\prime \|^2_{T^{-1}} \right\}\\
\label{pcpredictor}
\bar{x}^{(n+1)} &=  2x^{(n+1)} -x^{(n)},
\end{align}
\end{linenomath}
where for a positive semidefinite matrix $A$ the norm $\|z\|_A$ is defined as $\sqrt{z^\top A z}$.

\subsubsection{The need to generalize the CP primal-dual framework}
\label{sec:cpgen}
To apply the CP primal-dual algorithm
to $Q$ for fixed $f_0$, we need to write Eq. (\ref{qup})
in the form of the objective function in Eq. (\ref{cpgen}).
Manipulating the expression for $Q$ and dropping all terms
that are constant with respect to $f$, we obtain
\begin{linenomath}
\begin{align}
\label{qtcc}
Q\left(c,f_0; f\right) &=
\frac{1}{2} f^\top K^\top ( D-E )
K f - f^\top K^\top b, \\
\notag
K &= \left(
\begin{array}{c}
K_1 \\
K_2
\end{array} \right)
=
\left(
\begin{array}{c}
A\left( f_0 \right )Z \\
Z
\end{array} \right), \\
\notag
D &= \left(
\begin{array}{cc}
D_1  & 0\\
0    & D_2
\end{array} \right), \;\;
E = \left(
\begin{array}{cc}
E_1  & 0\\
0    & 0
\end{array} \right), \;\;
b = \left(
\begin{array}{c}
b_1\\
b_2
\end{array} \right) \\
\notag
D_1 &= \begin{cases}
\diag \left[ \hat{c}\left(f_0\right) \right] &\text{if } L = L_\text{TPL} \\
\mathbf{I} &\text{if } L = L_\text{LSQ}
\end{cases},\\
\notag
D_2 &= \begin{cases}
\diag \left[ A\left(f_0\right)^\top r_-\left(f_0\right) \right] &\text{if } L = L_\text{TPL} \\
\diag \left[ A\left(f_0\right)^\top r^\text{(log)}_-\left(f_0\right) \right] &\text{if } L = L_\text{LSQ}
\end{cases},\\
\notag
E_1 &= \begin{cases}
\diag \left[ r_-\left(f_0\right) \right] &\text{if } L = L_\text{TPL} \\
\diag \left[ r^\text{(log)}_-\left(f_0\right) \right] &\text{if } L = L_\text{LSQ}
\end{cases},\\
\notag
b_1 &=\begin{cases}
(D_1 - E_1) K_1 f_0 - r\left(f_0\right) &\text{if } L = L_\text{TPL} \\
(D_1 - E_1) K_1 f_0 - r^\text{(log)}\left(f_0\right)&\text{if } L = L_\text{LSQ}
\end{cases},\\
\notag
b_2 &= D_2 K_2 f_0.
\end{align}
\end{linenomath}
The matrices $D$ and $E$ are nonnegative and depend on $c$ and $f_0$;
$b$ is a vector which also depends on $c$ and $f_0$; and $K$ is a matrix that
depends only on $f_0$. Both terms of
$Q$ are functions of $K f$ and accordingly $Q$ is
identified with the function $F$ in the objective function of Eq. (\ref{cpgen})
\begin{linenomath}
\begin{align}
Q\left(c,f_0; f\right) &= F_Q(Kf), \notag \\
\label{FQ}
F_Q(z) &= \frac{1}{2} z^\top (D-E) z -z^\top b.
\end{align}
\end{linenomath}
Because $Q$ is a convex function of $f$, $F_Q$ is convex as a function of $f$.
The function $F_Q$, however, is {\it not} a convex function of $z$.
Because $D$ and $E$ are non-negative matrices, $F_Q$ is a difference
of convex functions of $z$,
\begin{linenomath}
\begin{align}
\notag
F_Q(z) & = F_{Q+}(z) - F_{Q-}(z), \\
\notag
F_{Q+}(z) & = \frac{1}{2} z^\top D z -z^\top b, \\
\notag
F_{Q-}(z) & = \frac{1}{2} z^\top E z,
\end{align}
\end{linenomath}
where $F_{Q+}(z)$ and $F_{Q-}(z)$ are convex functions of $z$.
That $F_Q(z)$ is not convex implies that $F_Q$ cannot be written
as the convex conjugate of $F_Q^*(y)$, and performing the maximization
over $y$ in Eq. (\ref{saddlept}) no longer yields Eq. (\ref{cpgen}).

\subsubsection{Heuristic derivation of MOCCA}
To generalize the CP algorithm to allow the case of interest,
we consider the function $F$ to be a convex-concave
\begin{linenomath}
\begin{equation}
\notag
F(z) = F_+(z) - F_-(z),
\end{equation}
\end{linenomath}
where $F_+$ and $F_-$ are both convex.
The heuristic strategy for MOCCA is to employ a convex approximation
to $F(z)$ in the neighborhood of a point $z=z_0$
\begin{linenomath}
\begin{equation}
\label{Fconvex}
F_\text{convex}(z_0;z) = F_+(z) - z^\top \nabla F_-(z_0);
\end{equation}
\end{linenomath}
(again we drop terms that are constant with respect to $z$). We then
execute an iteration
of the CP algorithm on the convex function $F_\text{convex}(z_0; z)$; and then
modify the point of convex approximation $z_0$ 
and repeat the iteration. The question then is how to choose
$z_0$, the center for the convex approximation, in light
of the fact that the optimization of $F$ in the CP algorithm
happens in the dual space with $F^*$, see Eq. (\ref{dualstep}).

A corresponding
primal point to a point in the dual space can be determined by
selecting the maximizer of the objective function in the definition
of the Legendre transform.
Taking the gradient of the objective function in Eq. (\ref{doublestar})
and setting it to zero,
yields
\begin{linenomath}
\begin{equation}
\label{dualcorrespondence}
x = \nabla_y F^*(y).
\end{equation}
\end{linenomath}
We use this relation to find the expansion point for the primal
objective function that mirrors
the current value of the dual variables.

Incorporating the convex approximation $F_\text{convex}(z_0; z)$ about the mirrored expansion point
$z_0$ into the CP algorithm, yields the iteration steps for MOCCA
\begin{linenomath}
\begin{align}
\label{mirrorpt}
z^{(n+1)}_0 &= \nabla_y F_\text{convex}^* \left(z^{(n)}_0;y^{(n)}\right)
= \Sigma^{-1} (y^{(n-1)} - y^{(n)} +\Sigma K \bar{f}^{(n-1)} ) \\
\label{moccadualstep}
y^{(n+1)} & =  \argmin_{y^\prime} \left\{F_\text{convex}^*\left(z^{(n+1)}_0; y^\prime\right)
+ \frac{1}{2} \|y^{(n)}+\Sigma K \bar{f}^{(n)} - y^\prime \|^2_{\Sigma^{-1}}
\right\}\\
\label{moccaprimalstep}
f^{(n+1)} & =  \argmin_{f^\prime} \left\{G(f^\prime)
+ \frac{1}{2} \|f^{(n)}-T K^\top y^{(n+1)} - f^\prime \|^2_{T^{-1}} \right\}\\
\label{moccapredictor}
\bar{f}^{(n+1)} &=  2f^{(n+1)} -f^{(n)},
\end{align}
\end{linenomath}
where $F_\text{convex}^*(z_0;y)$ is convex conjugate to $F_\text{convex}(z_0; z)$ 
with respect to the second 
argument; the first line
obtains the mirror expansion point using Eq. (\ref{dualcorrespondence})
and the right hand side expression is found by setting to zero the gradient of
the objective function in Eq. (\ref{moccadualstep});
the second line makes use of convex approximation $F_\text{convex}$ in the form
of its convex conjugate; and the remaining two lines are the same
as the those of the CP algorithm. For the simulations in this article,
all variables are initialized to zero. 
Convergence of MOCCA, the algorithm specified by
Eqs. (\ref{mirrorpt}) - (\ref{moccapredictor}), is investigated in an
accompanying paper \cite{Barber2015}, which also develops the algorithm for a more
general setting.

\subsubsection{Application of MOCCA to optimization of the spectral CT data fidelity}
\label{sec:onestepMOCCA}
The MOCCA algorithm handles a fixed convex-concave function $F$, 
convex function $G$, and linear transform $K$. In order to apply it to the
spectral CT data fidelity, we propose: employing the local quadratic expansion
in Eq. (\ref{qtcc}) to which we apply MOCCA, re-expand the spectral CT data discrepancy
at the current estimate of the material maps, and iterate this procedure until convergence.
We refer to iterations of the core MOCCA algorithm as ``inner'' iterations, and
the process of iteratively re-expanding the data discrepancy and applying MOCCA are the ``outer'' iterations.
Because MOCCA allows for non-smooth terms,
the convex constraints described in Sec. \ref{sec:constrainedOpt} can be incorporated
and the inner iterations aim at solving the intermediate problem
\begin{linenomath}
\begin{equation}
\label{innerOpt}
f^* = \argmin_f \left\{ \sum_{w \ell}
Q\left(c_{w \ell},f_0; f\right) + \sum_i \delta(P_i) \right\}.
\end{equation}
\end{linenomath}
For the remainder of this section, for brevity, we drop the constraints and write the update
steps taking only for the spectral CT data fidelity. The full algorithm with the convex
constraints discussed in Sec. \ref{sec:constrainedOpt} can be derived using the methods
described in \cite{Sidky2012} and an algorithm instance with TV constraints on the material
maps is covered in Appendix \ref{sec:onestepalgorithm}.

In applying MOCCA to $Q\left(c_{w \ell},f_0; f\right)$, we use the convex and concave
components from $F_Q$ in Eq. (\ref{FQ}) to form the local convex quadratic expansion needed in MOCCA,
see Eq. (\ref{Fconvex}),
\begin{linenomath}
\begin{equation}
\label{FQconvex}
F_{Q,\text{convex}}(z) = \frac{1}{2} z^\top D z - (z-z_0)^\top (b+E z_0).
\end{equation}
\end{linenomath}
The corresponding dual function
\begin{linenomath}
\begin{equation}
\label{FQconvex-dual}
F_{Q,\text{convex}}^*(y) = \frac{1}{2 D} \| y+b +E z_0 \|_2^2 + z^\top_0 E z_0,
\end{equation}
\end{linenomath}
is needed to derive the MOCCA dual update step at Eq. (\ref{moccadualstep}).
We note that because the material maps $f$ enter $Q\left(c_{w \ell},f_0; f\right)$
only after linear transformation, $Kf$, and comparing with the generic optimization problem
in Eq. (\ref{cpgen}),
we have $G(f) = 0$ for the present case where we only consider minimization of the data
discrepancy.

In using an inner/outer iteration, a basic question is how accurately does the inner
problem need to be solved. It turns out that it is sufficient to employ a {\it single}
inner iteration, so that effectively the proposed algorithm no longer consists of nested
iteration loops. Instead, the proposed algorithm performs re-expansion at every iteration:
\begin{linenomath}
\begin{align}
\label{sct_f0}
f_0 &= \bar{f}^{(n)} \\
\label{sct-stepsizes}
\Sigma^{(n)} &= |K_1(f_0)|\mathbf{1}/\lambda ; \; \;
T^{(n)} = \lambda  |K_1(f_0)|^\top \mathbf{1} \\
\label{sct-mirrorpt}
z^{(n+1)}_0 &=
(\Sigma^{(n)} )^{-1} \left(y^{(n-1)} - y^{(n)} +\Sigma^{(n)}  K_1(f_0) \bar{f}^{(n-1)} \right) \\
\label{sct-moccadualstep}
y^{(n+1)} & = (D_1(f_0) + \Sigma^{(n)} )^{-1} \left[ D_1(f_0)
\left( y^{(n)} + \Sigma^{(n)}  K_1(f_0) \bar{f}^{(n)} \right) -
 \Sigma^{(n)} \left( b_1(f_0) + E_1(f_0) z^{(n+1)}_0 \right) \right] \\
\label{sct-moccaprimalstep}
f^{(n+1)} & =  \bar{f}^{(n)} - T^{(n)} K_1(f_0)^\top y^{(n+1)} \\
\label{sct-moccapredictor}
\bar{f}^{(n+1)} &=  2f^{(n+1)} -f^{(n)},
\end{align}
\end{linenomath}
where $f^{(0)}$, $f^{(-1)}$, $\bar{f}^{(0)}$, $y^{(0)}$, and $y^{(-1)}$ are initialized
to zero vectors.

Before explaining each line of the one-step spectral CT algorithm specified by
Eqs. (\ref{sct_f0})-(\ref{sct-moccapredictor}),
we point out important features of the use of re-expansion at every iteration:
(1) There are no nested loops.
(2) The size of the system of equations is significantly
reduced; note that only the first matrix block of $K$, $D$, $E$, and $b$ (see Eq. (\ref{qtcc})
for their definition) appears in the steps of the algorithm. By re-expanding at every iteration
the set of update steps for the second matrix block becomes trivial.
(3) Re-expanding at every step is not guaranteed to converge, and
an algorithm control parameter $\lambda$ is introduced that balances algorithm convergence
rate against possible unstable iteration, see Sec. \ref{sec:results}
for a demonstration on how $\lambda$ impacts convergence.
A similar strategy was used together with the
CP algorithm in the use of non-convex image regularity norms, see \cite{Sidky2014}.

The first line of the algorithm, Eq. (\ref{sct_f0}), explicitly assigns the current material maps
estimate to the new expansion point. In this way it is clear in the following steps whether
$\bar{f}^{(n)}$ enters the equations through the re-expansion center or through the steps of MOCCA.
For the spectral CT algorithm it is convenient to use the vector step-sizes $\Sigma^{(n)}$
and $T^{(n)}$, defined in Eq. (\ref{sct-stepsizes}),
from the pre-conditioned form of the CP algorithm \cite{Pock2011}, because
the linear transform $ K_1(f_0)$ is changing at each iteration as the expansion center changes.
Computation of the vector step-sizes only involves single matrix-vector products of $|K_1(f_0)|$
and $|K_1(f_0)|^\top$ with a vector of ones, $\mathbf{1}$,
as opposed to performing the power method on $K_1(f_0)$ to find the scalar
step-sizes $\sigma$ and $\tau$, which would render re-expansion at every iteration impractical.
In Eq. (\ref{sct-stepsizes}), the parameter $\lambda$ enters in such a way that the
product $\| \Sigma^{1/2} K T^{1/2} \|$ remains constant.
For the preconditioned CP algorithm, $\lambda$ defined in
this way will not violate the step-size condition.
The dual and primal steps in Eqs. (\ref{sct-moccadualstep}) and (\ref{sct-moccaprimalstep}), respectively,
are obtained by analytic computation of the minimizations in Eqs. (\ref{moccadualstep}) and
(\ref{moccaprimalstep}) using Eq. (\ref{FQconvex}) and $G(f) = 0$, respectively.
The primal step at Eq.(\ref{sct-moccaprimalstep}) and the primal
variable prediction step at Eq. (\ref{sct-moccapredictor}) are identical to the corresponding
CP algorithm steps at Eqs. (\ref{primalstep}) and (\ref{predictor}), respectively.
The presented algorithm accounts only for the spectral CT data fidelity optimization. For
the full algorithm incorporating TV constraints used in the results section, see the pseudocode
in Appendix \ref{sec:onestepalgorithm}.

\subsection{One-step algorithm $\mu$-preconditioning}
\label{sec:preconditioning}

One of the main challenges of spectral CT image reconstruction is the similar
dependence of the linear X-ray attenuation curves on energy for different tissues/materials.
This causes rows of the attenuation matrix $\mu_{mi}$ to be nearly linearly dependent,
or equivalently its condition number is large. There are two effects of the poor conditioning
of $\mu_{mi}$: (1) the ability to separate the material maps is highly sensitive to inconsistency
in the spectral CT transmission data, and (2) the poor conditioning of $\mu_{mi}$ contributes
to the overall poor conditioning of spectral CT image reconstruction negatively impacting
algorithm efficiency. To address the latter issue, we introduce a simple preconditioning
step that orthogonalizes the attenuation curves. We call this step ``$\mu$''-preconditioning
to differentiate it from the preconditioning of the CP algorithm.

To perform $\mu$-preconditioning, we form the matrix
\begin{linenomath}
\begin{equation}
\label{musym}
M_{m m^\prime} = \sum_i \mu_{mi} \mu^\top_{im^\prime},
\end{equation}
\end{linenomath}
and perform the eigenvalue decomposition
\begin{linenomath}
\begin{equation}
\notag
M = U \diag(s) U^T,
\end{equation}
\end{linenomath}
where the eigenvalues are ordered $s_1 \ge s_2 \ge \dots \ge  s_{N_m}$.
The singular values of $\mu$ are given by the $\sqrt{s_i}$'s and its condition number
is $\sqrt{s_1/s_{N_m}}$. The preconditioning matrix for $\mu$ is given
by
\begin{linenomath}
\begin{align}
\label{mupc}
P & = \diag (\sqrt{s}) U^T, \\
\notag
P^{-1} &= U \diag (1/ \sqrt{s}).
\end{align}
\end{linenomath}
Implementation of $\mu$-preconditioning consists of the
following steps:
\begin{itemize}
\itemsep1em
\item \textbf{Transformation of material maps and attenuation matrix} -
The appropriate transformation is arrived at through inserting the identity
matrix in the form of $P^{-1} P$ into the exponent of
the intensity counts data model
in Eq. (\ref{meanModel}):
\begin{linenomath}
\begin{equation}
\label{meanModelExp}
\sum_{m k}  X_{\ell k} \mu_{m i} f_{k m}
= \sum_{m\, m^\prime m^{\prime\prime} k} 
X_{\ell k} \mu_{m^{\prime\prime} i} (P^{-1})_{m^{\prime\prime} m} P_{m \, m^\prime}f_{k m^\prime}
= \sum_{m k}  X_{\ell k}\mu^\prime_{m i} f^\prime_{k m},
\end{equation}
\end{linenomath}
where
\begin{linenomath}
\begin{align}
\label{ftrans}
f^\prime_{k m} &= \sum_{m^\prime} P_{m m^\prime}f_{k m^\prime}, \\
\label{mutrans}
\mu^\prime_{m i} &=  \sum_{m^\prime}\mu_{m^\prime i} (P^{-1})_{ m^{\prime}m}.
\end{align}
\end{linenomath}

\item \textbf{Substitution into the one-step algorithm} -
Substitution of the transformed material maps and attenuation matrix into
the one-step algorithm given by Eqs. (\ref{sct_f0})-(\ref{sct-moccapredictor})
is fairly straight-forward. All occurrences of $f$ are replaced by $f^\prime$,
and the linear transform $K_1$ is replaced by
\begin{linenomath}
\begin{equation}
\notag
K_1^\prime = A^\prime(f_0^\prime) Z^\prime,
\end{equation}
\end{linenomath}
where, using Eqs. (\ref{zdef}) and (\ref{adef}),
\begin{linenomath}
\begin{equation}
\notag
A^\prime_{w \ell, \ell^\prime i}(f^\prime) =  \frac{s_{w \ell i} \exp \left[- (Z^\prime f^\prime)_{\ell i} \right]}
{\sum_{i^\prime} s_{w \ell i^\prime} \exp \left[- (Z^\prime f^\prime)_{\ell i^\prime} \right]},
\end{equation}
\end{linenomath}
and
\begin{linenomath}
\begin{equation}
\notag
Z^\prime = \mu^\prime X.
\end{equation}
\end{linenomath}
Using $\mu$-preconditioning, care must be taken in computing the vector stepsizes
$\Sigma^\prime$ and $T^\prime$ in Eq. (\ref{sct-stepsizes}). Without $\mu$-preconditioning,
the absolute value symbols are superfluous, because $K_1$ has non-negative
matrix elements. With $\mu$-preconditioning, the absolute value operation is necessary,
because $K^\prime_1$ may have negative entries through its dependence on $Z^\prime$
and in turn $\mu^\prime$.
\item \textbf{Formulation of constraints} - the previously discussed constraints
are functions of the untransformed material maps. As a result, in using $\mu$-preconditioning
where we solve for the transformed material maps, the constraints should be formulated
in terms of 
\begin{linenomath}
\begin{equation}
\label{freverse}
f = P^{-1}f^\prime .
\end{equation}
\end{linenomath}
The explicit pseudocode for constrained data-discrepancy minimization using $\mu$-preconditioning
is given in Appendix \ref{sec:onestepalgorithm}.
\end{itemize}
After applying the $\mu$-preconditioned one-step
algorithm the final material maps are arrived at through Eq. (\ref{freverse}).


\subsection{Convergence checks}
Within the present primal-dual framework we employ the primal-dual gap for checking convergence.
The primal-dual gap that we seek is the difference between the convex quadratic
approximation using the first matrix block in Eq. (\ref{FQconvex}),
which is the objective function in the primal minimization
\begin{linenomath}
\begin{equation}
\label{FQconvex-min}
f^\star = \argmin_f \left\{ \frac{1}{2} (K_1 f)^\top D_1 K_1 f - (K_1 f-z_0)^\top (b_1+E_1 z_0)
\right\} ,
\end{equation}
\end{linenomath}
and the objective function in the Fenchel dual maximization problem
\begin{linenomath}
\begin{equation}
\label{FQconvex-Fenchel-dual}
y^\star =\argmax_y \left\{ -\frac{1}{2} D^{-1}_1\| y+b_1 +E_1 z_0 \|_2^2 - z^\top_0 E_1 z_0 \right\}
\text{ such that } K_1^\top y=0.
\end{equation}
\end{linenomath}
These problems are derived from the general forms in Eqs. (\ref{cpgen}) and (\ref{cpgen_dual}),
and the constraint in the dual maximization comes from the fact that $G(f)=0$ in the primal
problem, see Sec. 3.1 in \cite{Sidky2012}. For a convergence check we inspect the difference
between these two objective functions. Note that the constant term $z^\top_0 E_1 z_0$ cancels in
this subtraction and plays no role in the optimization algorithm, and could thus be left out.
If the material maps $f^{(n)}$ attain a stable value, the constraint $K^\top_1 y =0$ 
is necessarily satisfied from inspection of Eq. (\ref{sct-moccaprimalstep}).
When other constraints are included the estimates of the material maps should be checked
against these constraints and the primal-dual gap is modified.

\section{Results}
\label{sec:results}

\begin{figure}[!h]
\begin{minipage}[b]{\linewidth}
\centering
\centerline{\includegraphics[width=0.5\linewidth]{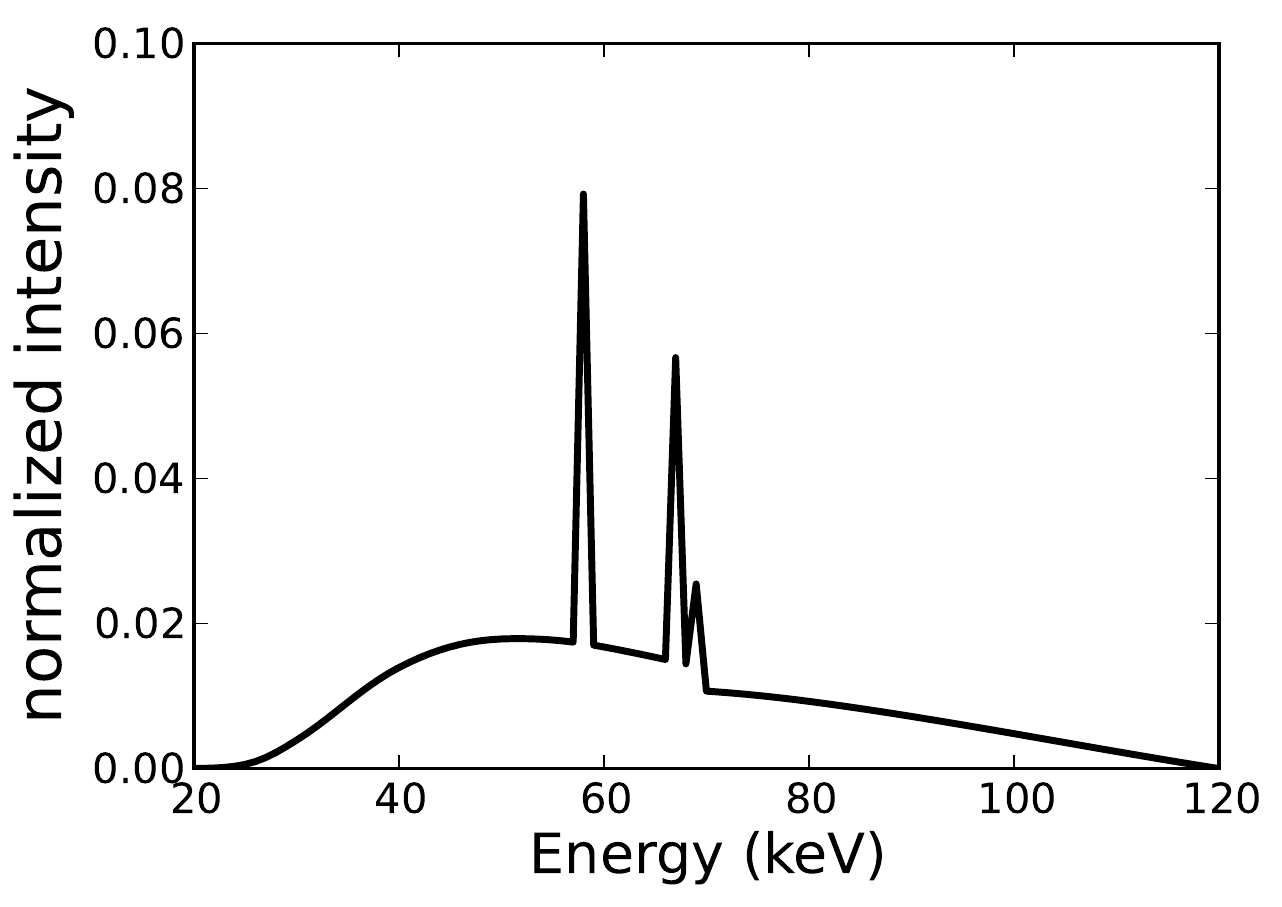}}
\end{minipage}
\caption{Normalized spectrum of a typical X-ray source for CT operating at a potential of 120kV.
\label{fig:spectrum}}
\end{figure}

We demonstrate use of the one-step spectral CT algorithm on simulated transmission data
modeling an ideal photon-counting detector. The X-ray spectrum, shown in Fig. \ref{fig:spectrum},
is assumed known. In modeling the ideal detector, the spectral response of an energy-windowed photon
count measurement is taken to be the same as that of Fig. \ref{fig:spectrum} between the bounding
threshold energies of the window and zero outside. We conduct two studies. The first is focused
on demonstrating convergence and application of the one-step algorithm
with recovery of material maps for a two-material head phantom using the following minimization problems
\begin{linenomath}
\begin{equation}
\label{TPL-TV}
\notag
\text{TPL-TV:} \; \; \; \argmin_{f} D_\text{TPL}(c,\hat{c}(f))
\text{ such that } \|f_m\|_\text{TV} \le \gamma_m  \; \forall \, m ,
\end{equation}
\end{linenomath}
and
\begin{linenomath}
\begin{equation}
\label{LSQ-TV}
\notag
\text{LSQ-TV:} \; \; \; \argmin_{f} D_\text{LSQ}(c,\hat{c}(f))
\text{ such that } \|f_m\|_\text{TV} \le \gamma_m \; \forall \, m .
\end{equation}
\end{linenomath}
The pseudo-code for TPL-TV and LSQ-TV is given explicitly in Appendix \ref{sec:onestepalgorithm}.
The second study simulates
a more realistic study demonstrating application
on an anthropomorphic chest phantom simulating multiple tissues/materials
at multiple densities. For this study we demonstrate one-step image reconstruction
of a mono-energetic image at energy $E$ using
\begin{linenomath}
\begin{equation}
\label{TPL-monoTV}
\notag
\text{TPL-monoTV:} \; \; \; \argmin_{f} D_\text{TPL}(c,\hat{c}(f))
\text{ such that } \|f^\text{(mono)}(E)\|_\text{TV} \le \gamma_\text{mono}.
\end{equation}
\end{linenomath}
Note that for monoenergetic image reconstruction, the TV constraint is placed on the
monoenergetic image, but the optimization is performed over the individual material maps $f_m$
and the monoenergetic image is formed after the optimization using Eq. (\ref{monoimage}).

Aside from the system specification parameters, such as number of views, detector bins,
and image dimensions, the algorithm parameters are the TV constraints $\gamma_m$ for
TPL-TV and LSQ-TV or $\gamma_\text{mono}$ for TPL-monoTV and the primal-dual step size
ratio $\lambda$. The TV constraints $\gamma_m$ or $\gamma_\text{mono}$ affect the image
regularization, but $\lambda$ is a tuning parameter which does not alter the solution
of TPL-TV, LSQ-TV, or TPL-monoTV. It is used to optimize convergence speed
of the one-step image reconstruction algorithm.

\subsection{Head phantom studies with material map TV-constraints}
\label{sec:head}

\begin{figure}[!h]
\begin{minipage}[b]{\linewidth}
\centering
\centerline{\includegraphics[width=0.6\linewidth]{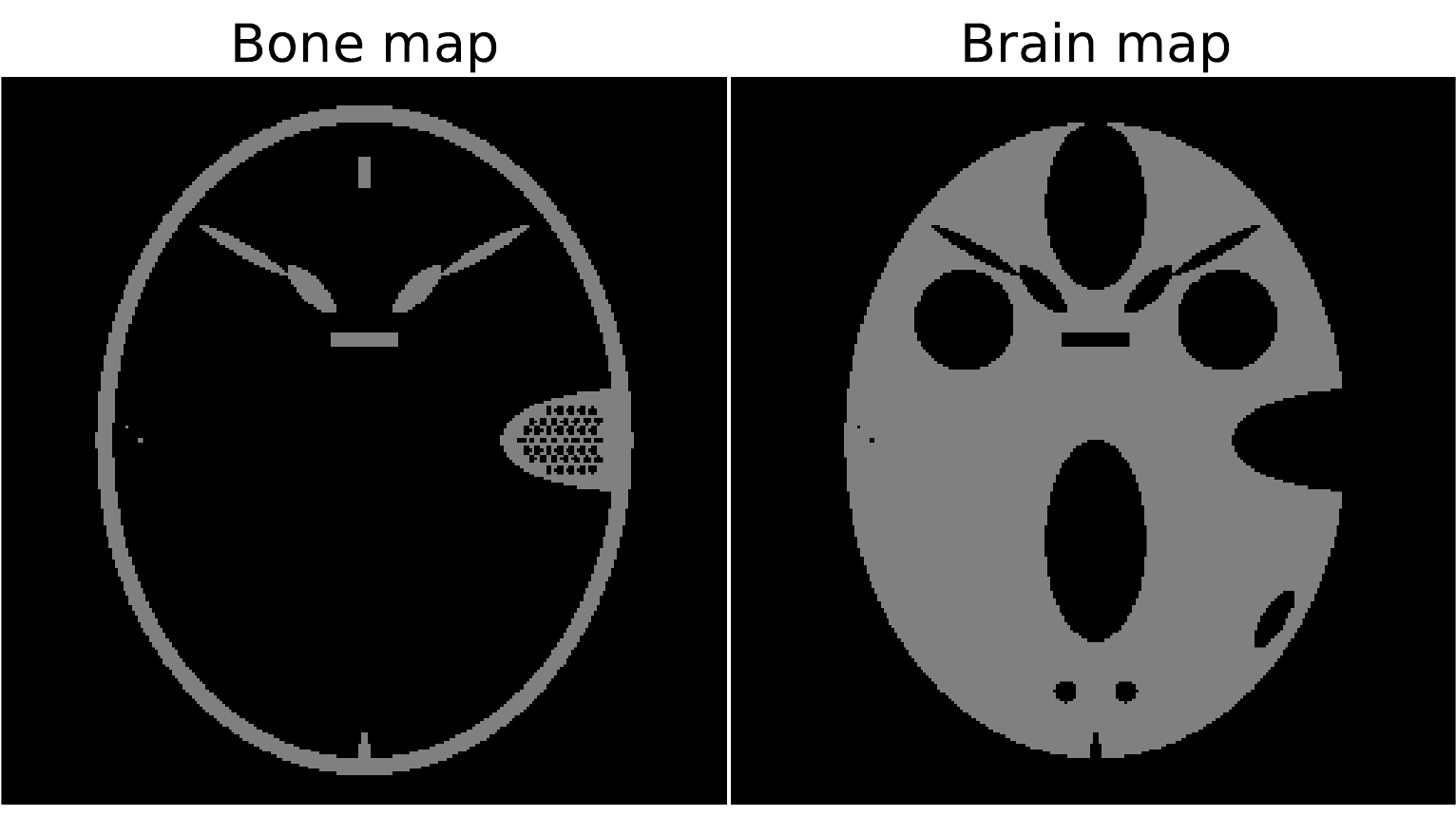}}
\end{minipage}
\caption{Bone and brain maps derived from the FORBILD head phantom.
Both images are shown in the gray scale window [0.9, 1.1].
\label{fig:FORBILD}}
\end{figure}

For the present studies, we employ a two-material phantom derived from the FORBILD head phantom
shown in Fig. \ref{fig:FORBILD}.  The spectral CT transmission counts are computed by use
of the discrete-to-discrete model in Eq. (\ref{meanModel}). The true material maps
$f_{k,\text{bone}}$ and $f_{k,\text{brain}}$ are the 256$\times$256 pixel arrays shown in
Fig. \ref{fig:FORBILD} and the corresponding linear X-ray coefficients $\mu_{\text{bone},i}$
and $\mu_{\text{brain},i}$ are obtained from the NIST tables available in Ref. \cite{Hubbell1995}
for energies ranging from 20 to 120 KeV in increments of 1 KeV.  By employing the same
data model as that used in the image reconstruction algorithm, we can investigate the
convergence properties of the one-step algorithm.

\begin{figure}[!h]
\begin{minipage}[b]{\linewidth}
\centering
\centerline{
\includegraphics[width=0.4\linewidth]{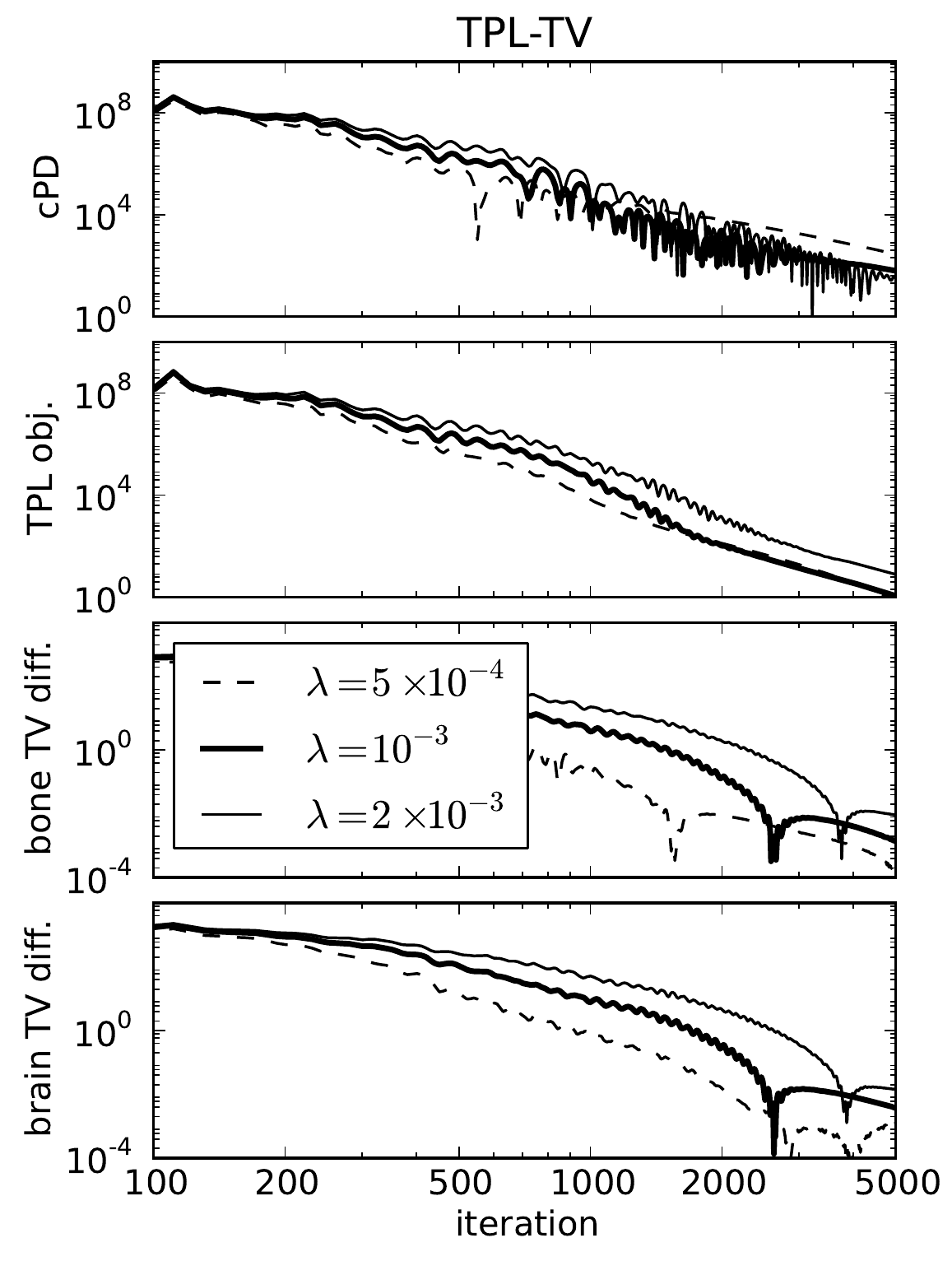}
\includegraphics[width=0.4\linewidth]{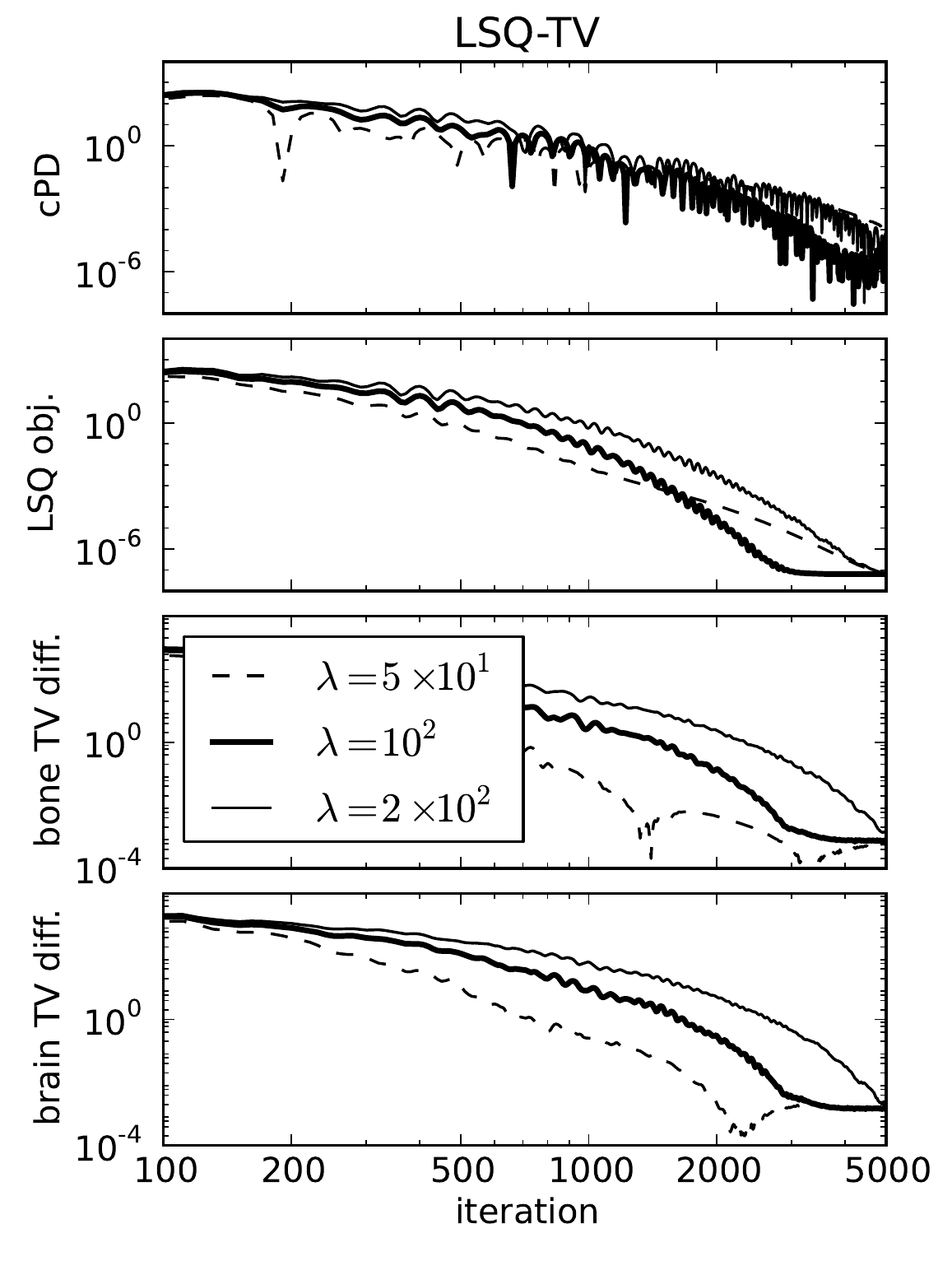}
}
\end{minipage}
\caption{Convergence metrics for LSQ-TV and TPL-TV and for different values of $\lambda$
with ideal, noiseless data.
First, second, third, and fourth rows show the conditional
primal-dual (cPD) gap, data discrepancy objective function, difference
between the TV of estimated bone map and that of the phantom bone map, and same
for the brain map TV. Note that the expressions for the gap and data discrepancy
are different for TPL and LSQ; thus those quantities are not directly comparable.
\label{fig:convergence}}
\end{figure}

For the head phantom simulations, the scanning configuration is 2D fan-beam CT with
a source to iso-center distance of 50 cm and source to detector distance of 100 cm. The physical
size of the phantom pixel array is $20 \times 20$ cm$^2$. The number of projection views
over a full 2$\pi$ scan is 128 and the number of detector bins along a linear detector array
is 512. This configuration is undersampled by a factor of 4 \cite{Jorgensen2013}. 
Two X-ray energy windows are simulated 
with a spectral response for each window given by the spectrum shown in Fig. \ref{fig:spectrum}
in the energy ranges [20 KeV, 70 KeV] and [70 KeV, 120 KeV] for the first and second energy
windows, respectively. 

\begin{figure}[!h]
\begin{minipage}[b]{\linewidth}
\centering
\centerline{
\includegraphics[width=0.4\linewidth]{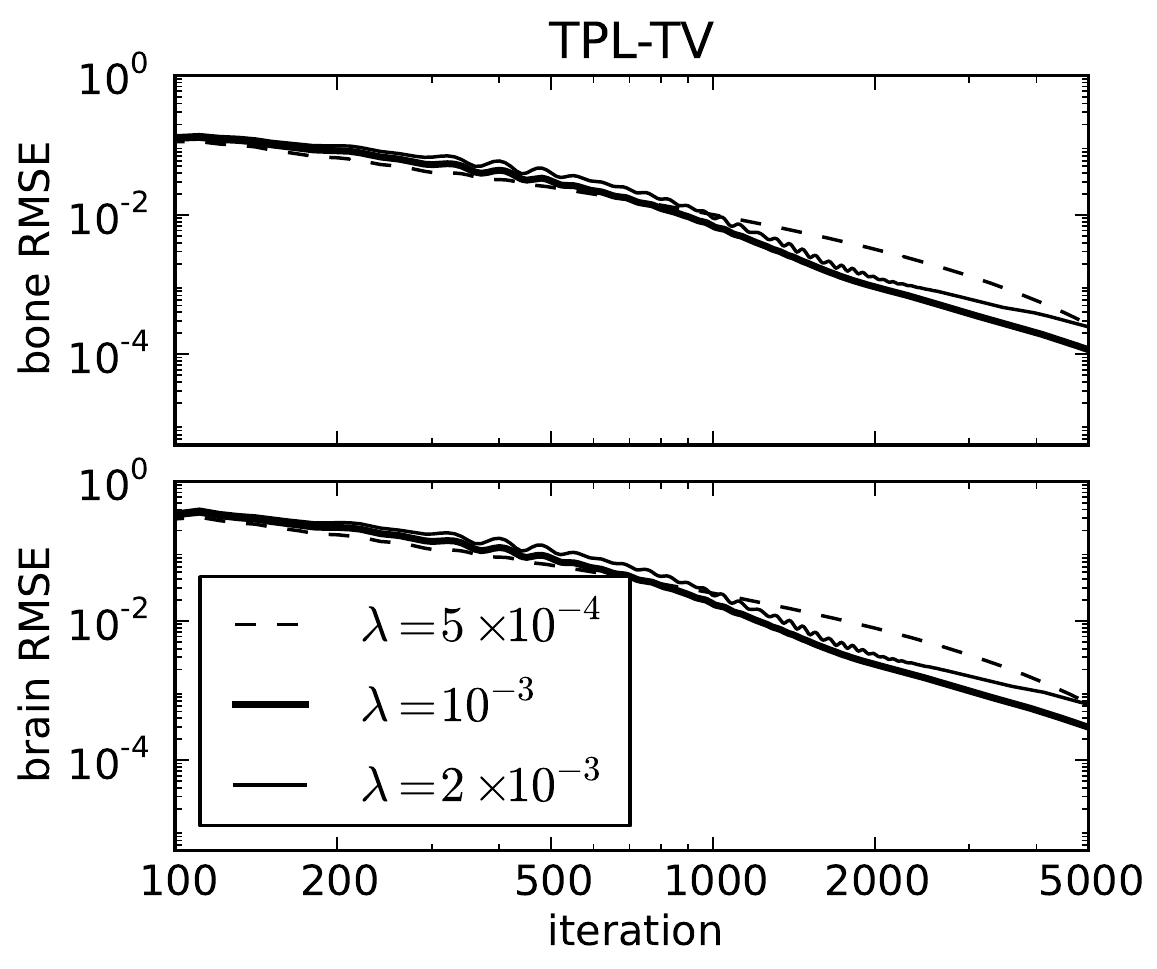}
\includegraphics[width=0.4\linewidth]{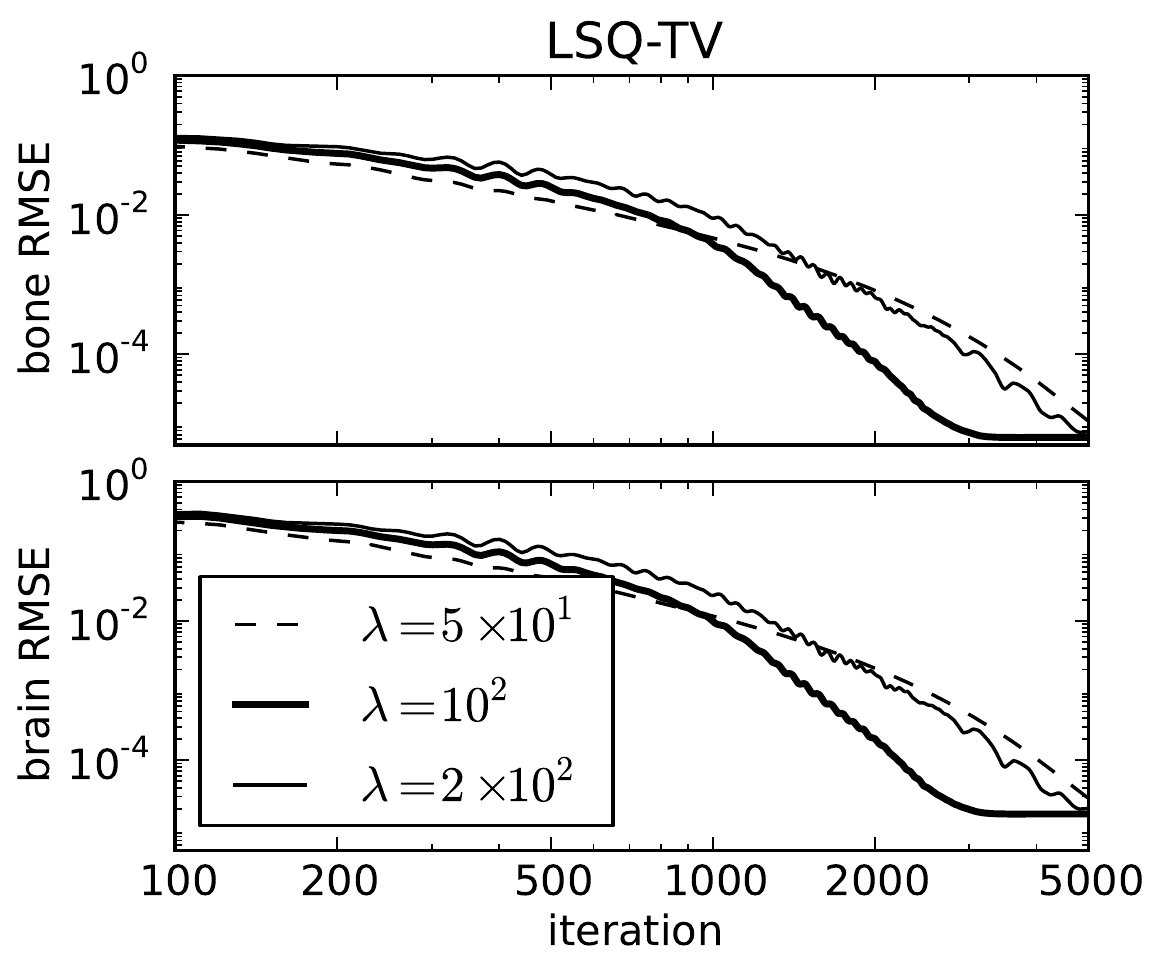}
}
\end{minipage}
\caption{Convergence of the material map estimates to the phantom material maps for
LSQ-TV and TPL-TV and for different values of $\lambda$ with ideal, noiseless data.
\label{fig:iconvergence}}
\end{figure}

\paragraph*{Ideal data study}
For ideal, noiseless data several image metrics are plotted in Fig. \ref{fig:convergence}
for different values of $\lambda$, and it is observed that the conditional primal-dual
(cPD) gap and data discrepancy
tend to zero while the material map TVs converge to the designed values. For this
problem the convergence metrics are the cPD and material map TVs; the data discrepancy
only tends to zero here due to the use of ideal data and in general when data inconsistency
is present the minimum data discrepancy will be greater than zero. The convergence
metrics demonstrate convergence of the one-step algorithm for the particular
problem under study. It is important, however, to inspect these metrics for each application
of the one-step algorithm, because there is no theoretical guarantee of convergence due
to the re-expansion step in Eq. (\ref{sct_f0}).
From the present results it is clear that progress towards convergence depends on $\lambda$;
thus it is important to perform a search over $\lambda$. 

\begin{figure}[!h]
\begin{minipage}[b]{\linewidth}
\centering
\centerline{
\includegraphics[width=0.45\linewidth]{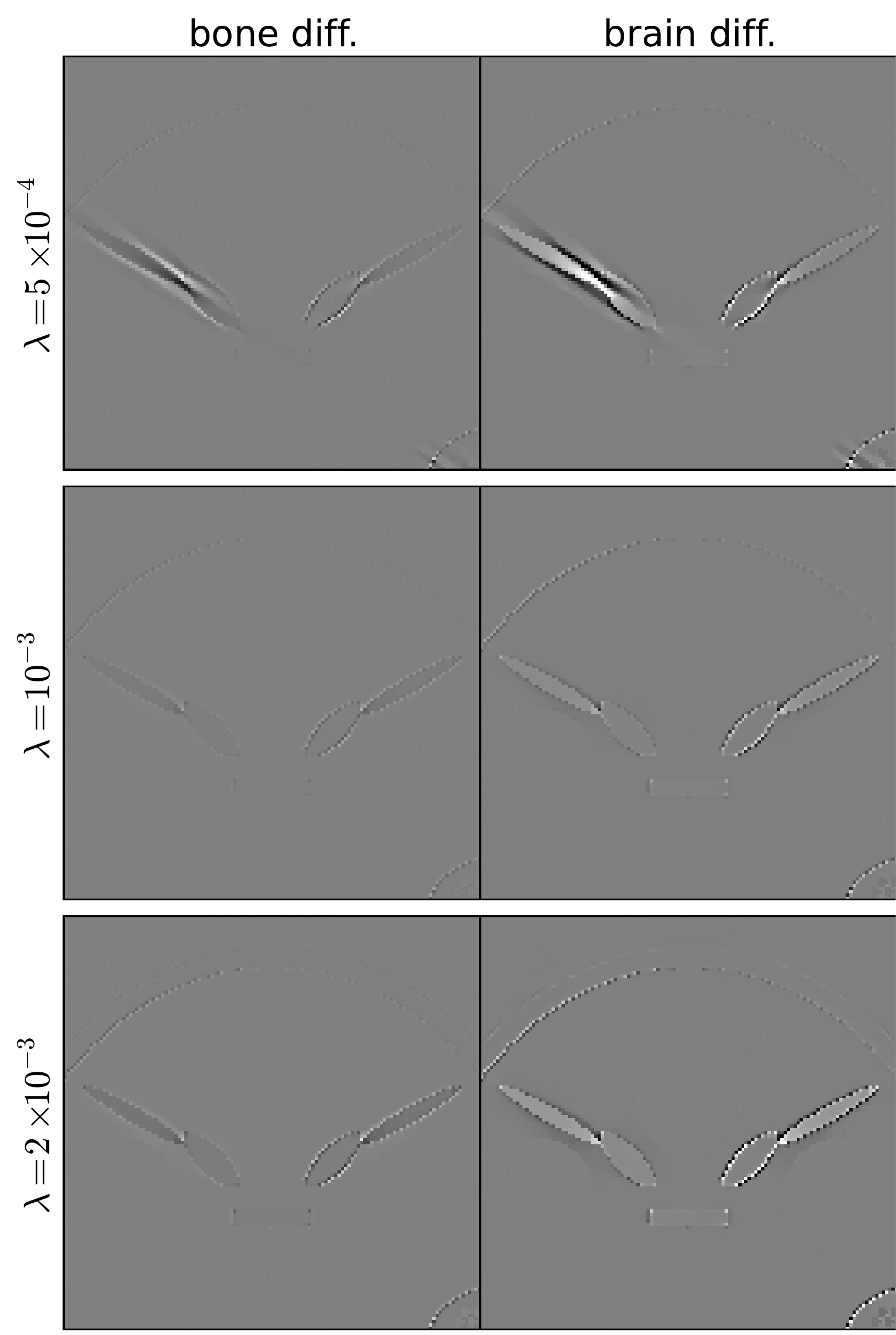}
\includegraphics[width=0.45\linewidth]{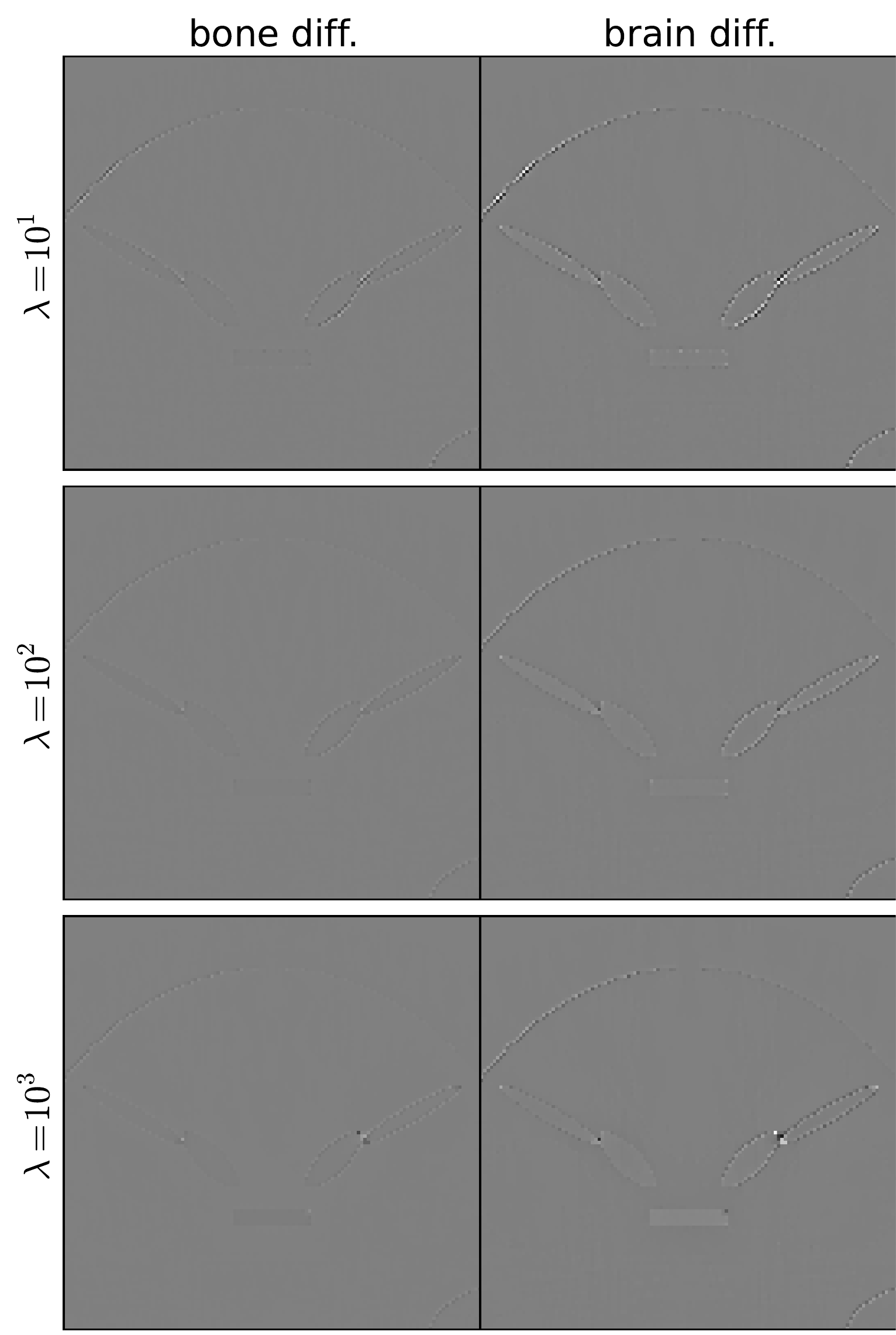}
}
\centerline{~~~TPL-TV~~~~~~~~~~~~~~~~~~~~~~~~~~~~~~~~~~~~~~~~~~~~~~~~~~~LSQ-TV}
\end{minipage}
\caption{Difference between estimated brain and bone maps after 5,000
iterations and the corresponding
phantom map shown in a 1\% gray scale window [-0.01, 0.01] for TPL-TV and
a 0.1\% window [-0.001, 0.001] for LSQ-TV
and different values of $\lambda$ with ideal, noiseless data.
The difference images are displayed in a region
of interest around the sinus bones.
\label{fig:idealdiff}}
\end{figure}

To demonstrate convergence of the material map estimates to the corresponding phantom,
we plot image root-mean-square-error (RMSE) in Fig. \ref{fig:iconvergence} 
as a function of iteration number and show
the map differences at the last iteration performed in Fig. \ref{fig:idealdiff}.
The material map estimates are seen to converge to the corresponding phantom maps
despite the projection view-angle under-sampling. Thus we note that the material
map TV constraints are effective at combatting these under-sampling artifacts
just as they are for standard CT \cite{Sidky2006,Sidky2008}. The image RMSE curves only
give a summary metric for material map convergence, and it is clear from the difference
images displayed in narrow gray scale window that convergence can be spatially non-uniform.
For these idealized examples the pre-conditioned one-step algorithm appears to be
more effective for LSQ-TV than TPL-TV as the image RMSE attained for the former is
significantly lower than that of the latter. In Fig. \ref{fig:iconvergence} curves for LSQ-TV
at $\lambda = 1 \times 10^2$, the image RMSE curves plateau at $~10^{-5}$ due to the
fact that the solution of LSQ-TV is achieved to the single precision accuracy of the computation.

\begin{figure}[!h]
\begin{minipage}[b]{\linewidth}
\centering
\centerline{
\includegraphics[width=0.7\linewidth]{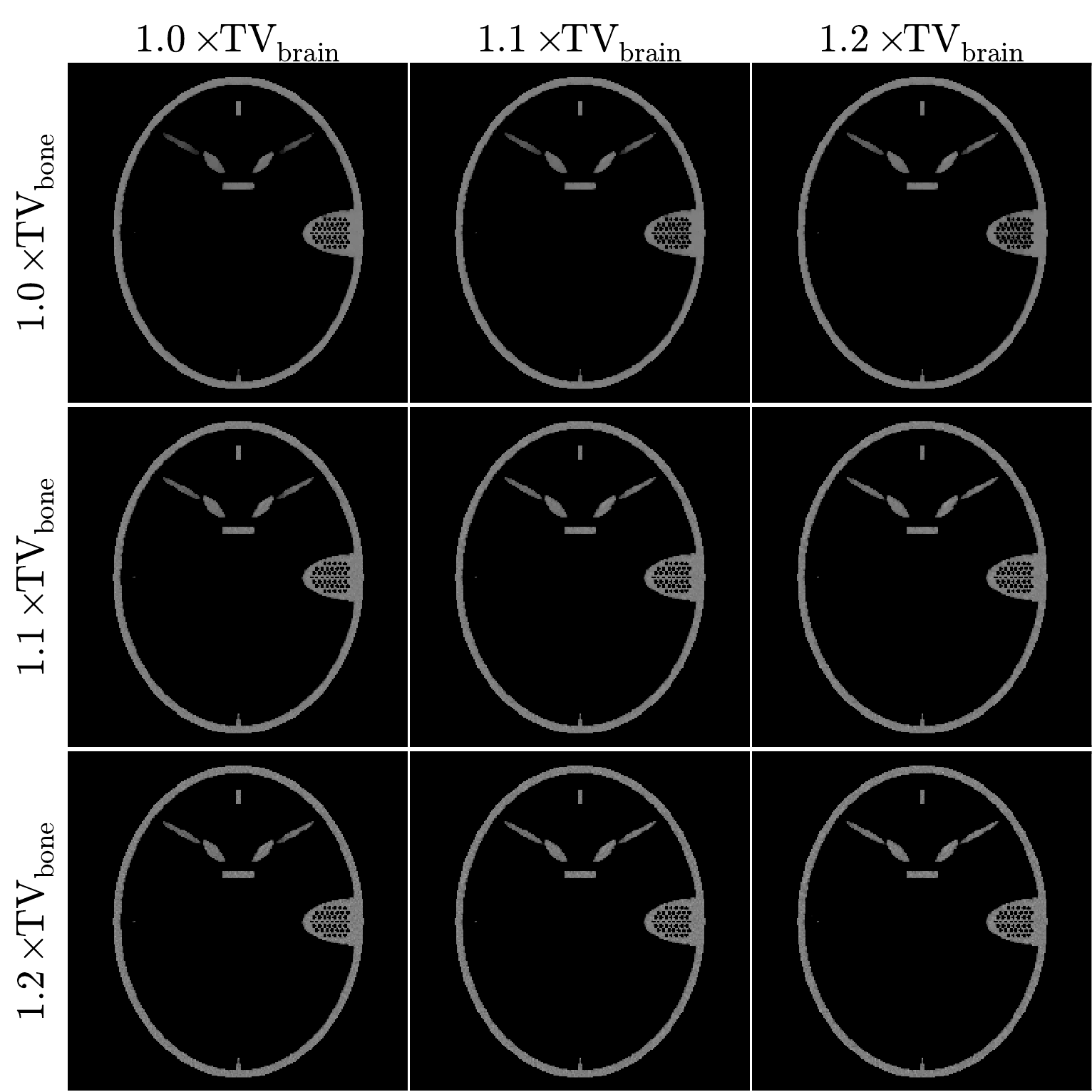}}
\end{minipage}
\caption{Reconstructed bone map by use of TPL-TV from simulated noisy projection
spectral CT transmission data. The material map TV constraints are varied according to fractions
of the corresponding phantom material map TV.
\label{fig:noisyTPLbone}}
\end{figure}

\begin{figure}[!h]
\begin{minipage}[b]{\linewidth}
\centering
\centerline{
\includegraphics[width=0.7\linewidth]{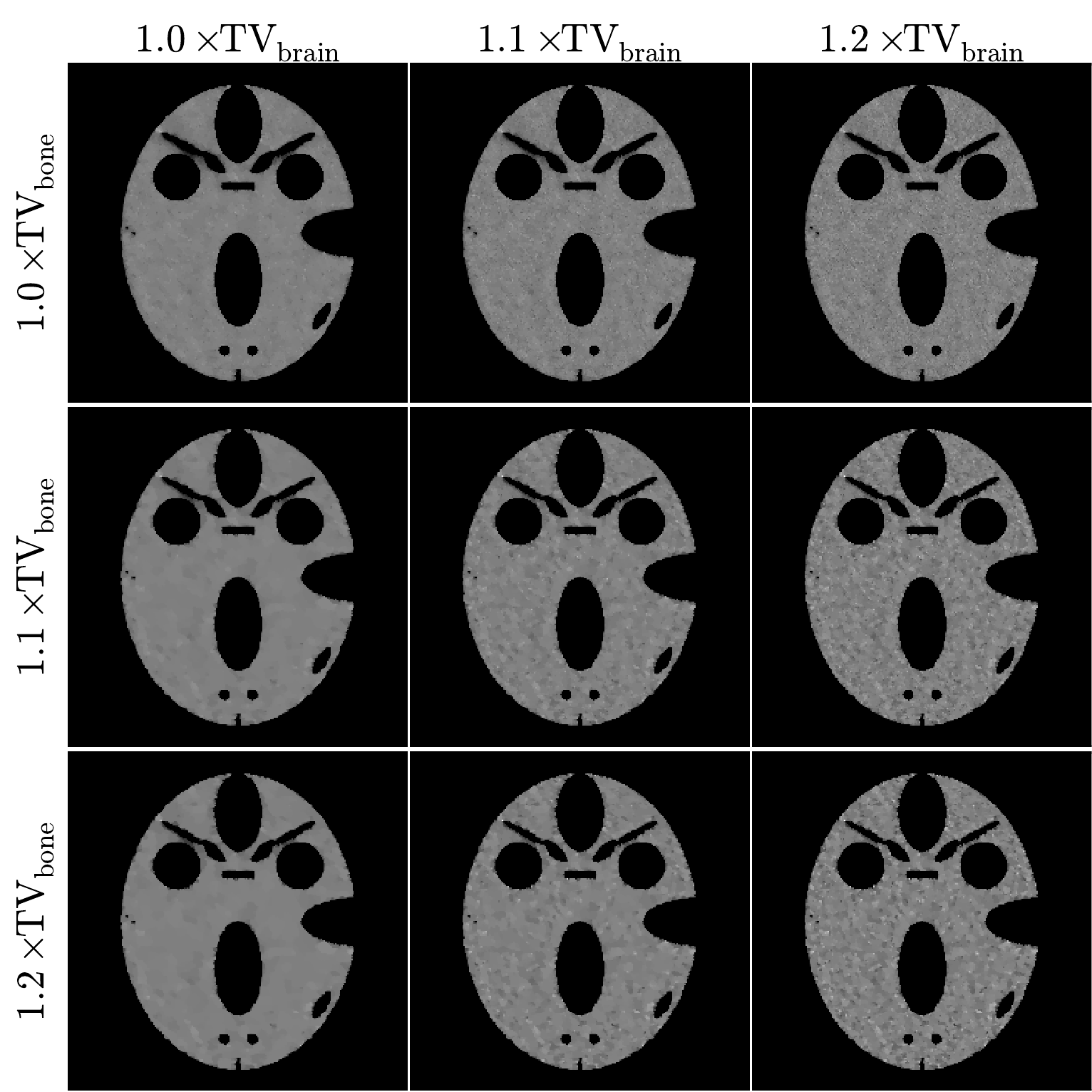}}
\end{minipage}
\caption{Reconstructed brain map by use of TPL-TV from simulated noisy projection
spectral CT transmission data. The material map TV constraints are varied according to fractions
of the corresponding phantom material map TV.
\label{fig:noisyTPLbrain}}
\end{figure}

\paragraph*{Noisy data study}

The noisy simulation parameters are identical to the previous noiseless study except
that the spectral CT data are generated from the transmission Poisson model.  The
mean of the transmission measurements is arrived at by assuming $4 \times 10^6$ total
photons are incident at each detector pixel over the complete X-ray spectrum. 
As the simulated scan acquires only 128 views, the total X-ray exposure is equivalent
to acquiring 512 views at $1 \times 10^6$ photons per detector pixel.

\begin{figure}[!h]
\begin{minipage}[b]{\linewidth}
\centering
\centerline{
\includegraphics[width=0.7\linewidth]{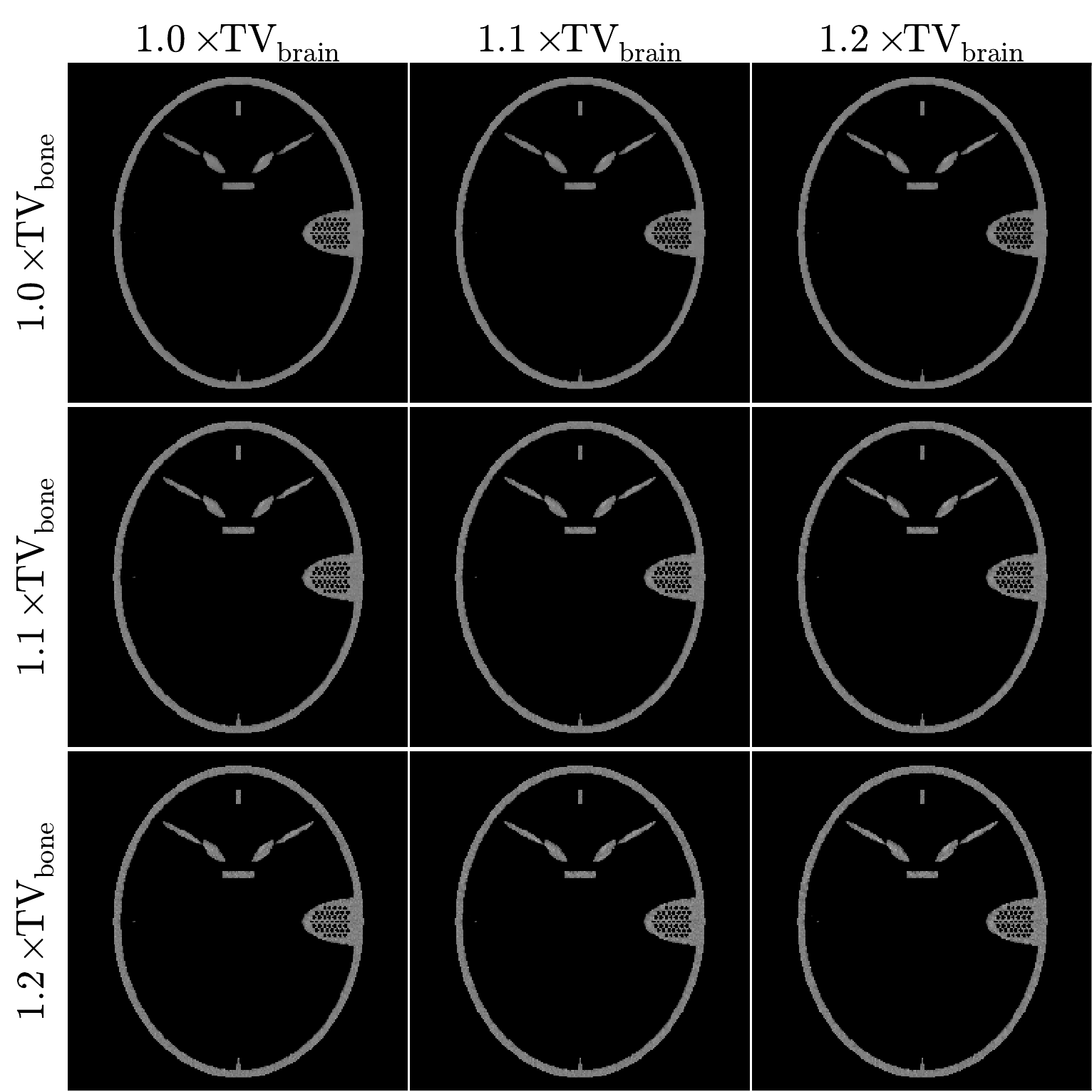}}
\end{minipage}
\caption{Reconstructed bone map by use of LSQ-TV from simulated noisy projection
spectral CT transmission data. The material map TV constraints are varied according to fractions
of the corresponding phantom material map TV.
\label{fig:noisyLSQbone}}
\end{figure}

\begin{figure}[!h]
\begin{minipage}[b]{\linewidth}
\centering
\centerline{
\includegraphics[width=0.7\linewidth]{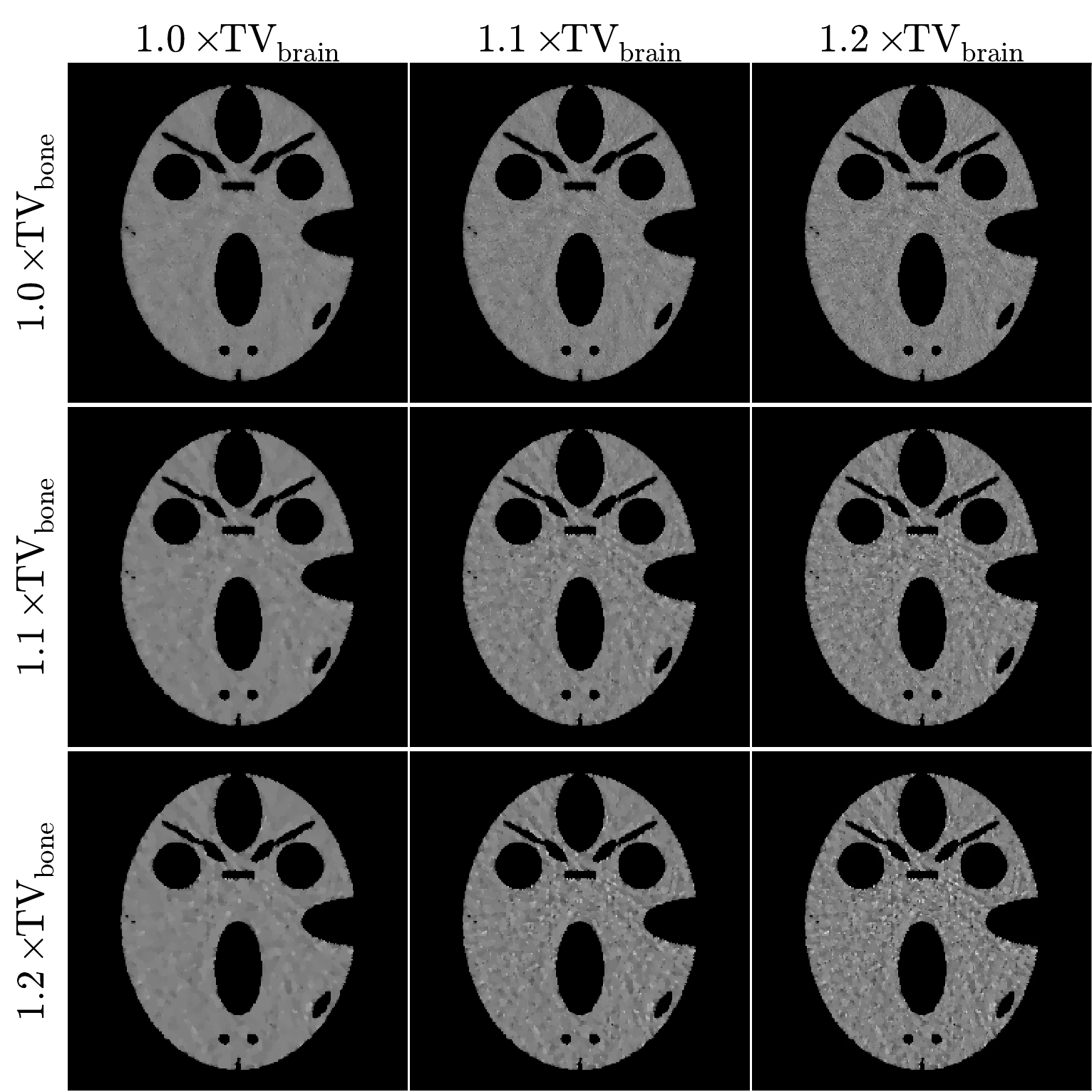}}
\end{minipage}
\caption{Reconstructed brain map by use of LSQ-TV from simulated noisy projection
spectral CT transmission data. The material map TV constraints are varied according to fractions
of the corresponding phantom material map TV.
\label{fig:noisyLSQbrain}}
\end{figure}

We obtain multiple material map reconstructions
varying the TV constraints among values greater than or equal
to the actual values of the known bone and brain maps.
The results for TV-TPL are shown in Figs. \ref{fig:noisyTPLbone} and \ref{fig:noisyTPLbrain},
and those for TV-LSQ are shown in Figs. \ref{fig:noisyLSQbone} and \ref{fig:noisyLSQbrain}.
In all images the bone and brain maps recover the main features of the true phantom maps,
and the main difference in the images is the structure of the noise. The noise texture 
of the recovered brain maps appears to be patchy for lower TV and grainy for larger TV
constraints. Also, in comparing the brain maps for TPL-TV in Fig. \ref{fig:noisyTPLbrain}
and LSQ-TV in Fig. \ref{fig:noisyLSQbrain}, streak artifacts are more apparent in the latter
particularly for the larger TV constraint values.

\begin{figure}[!h]
\begin{minipage}[b]{\linewidth}
\centering
\centerline{
\includegraphics[width=0.4\linewidth]{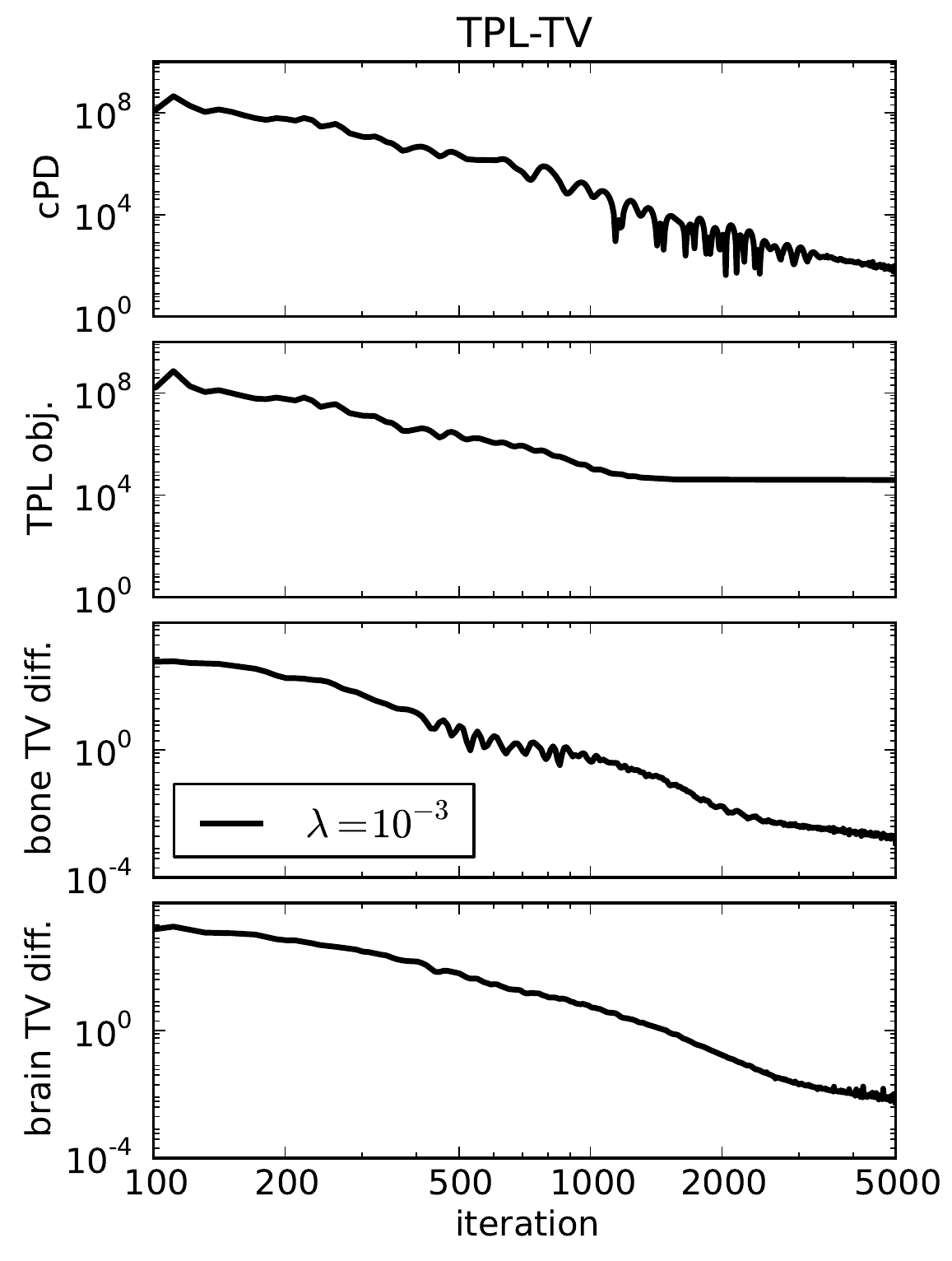}
\includegraphics[width=0.4\linewidth]{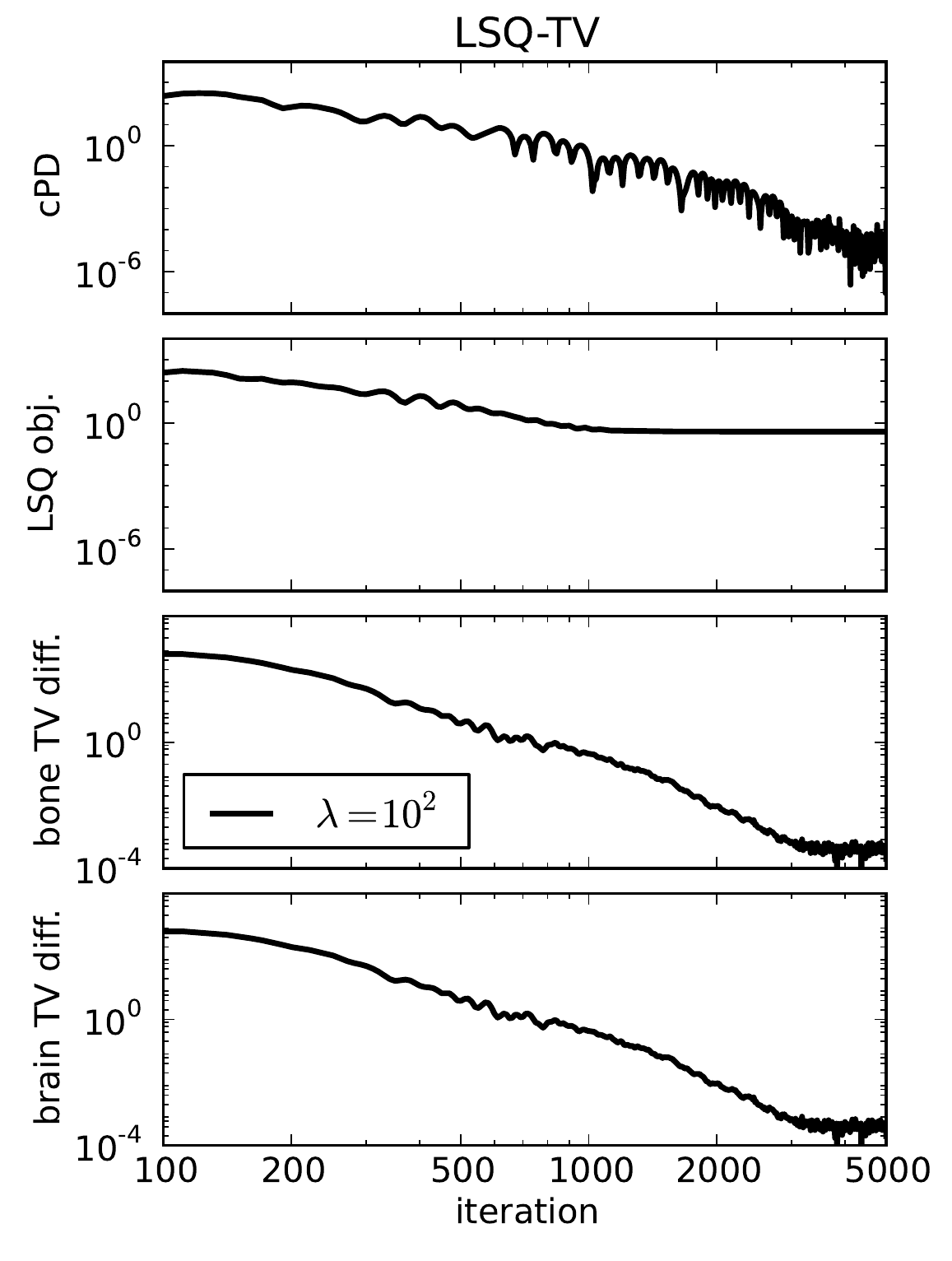}
}
\end{minipage}
\caption{Same as Fig.~\ref{fig:convergence}
except that only one value of $\lambda$ is shown and
the results are for noisy data and the TV constraints for
the bone and brain maps are
set to $1.1\times \text{TV}_\text{bone}$ and $1.1\times \text{TV}_\text{brain}$, respectively.
The TV constraint settings correspond to the center images in
Figs.~\ref{fig:noisyTPLbone}-\ref{fig:noisyLSQbrain}.
\label{fig:noisyConvergence}}
\end{figure}

\begin{figure}[!h]
\begin{minipage}[b]{\linewidth}
\centering
\centerline{
\includegraphics[width=0.4\linewidth]{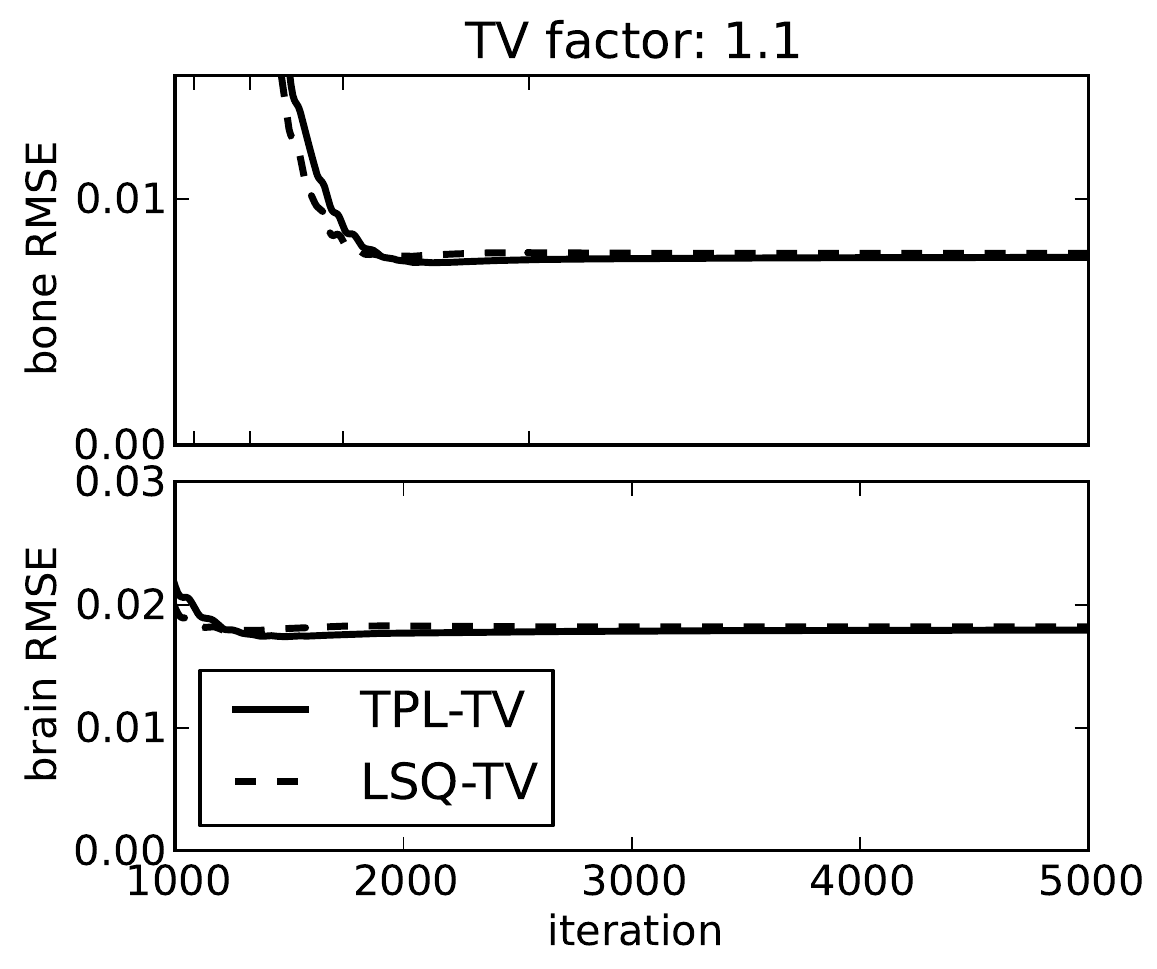}
\includegraphics[width=0.4\linewidth]{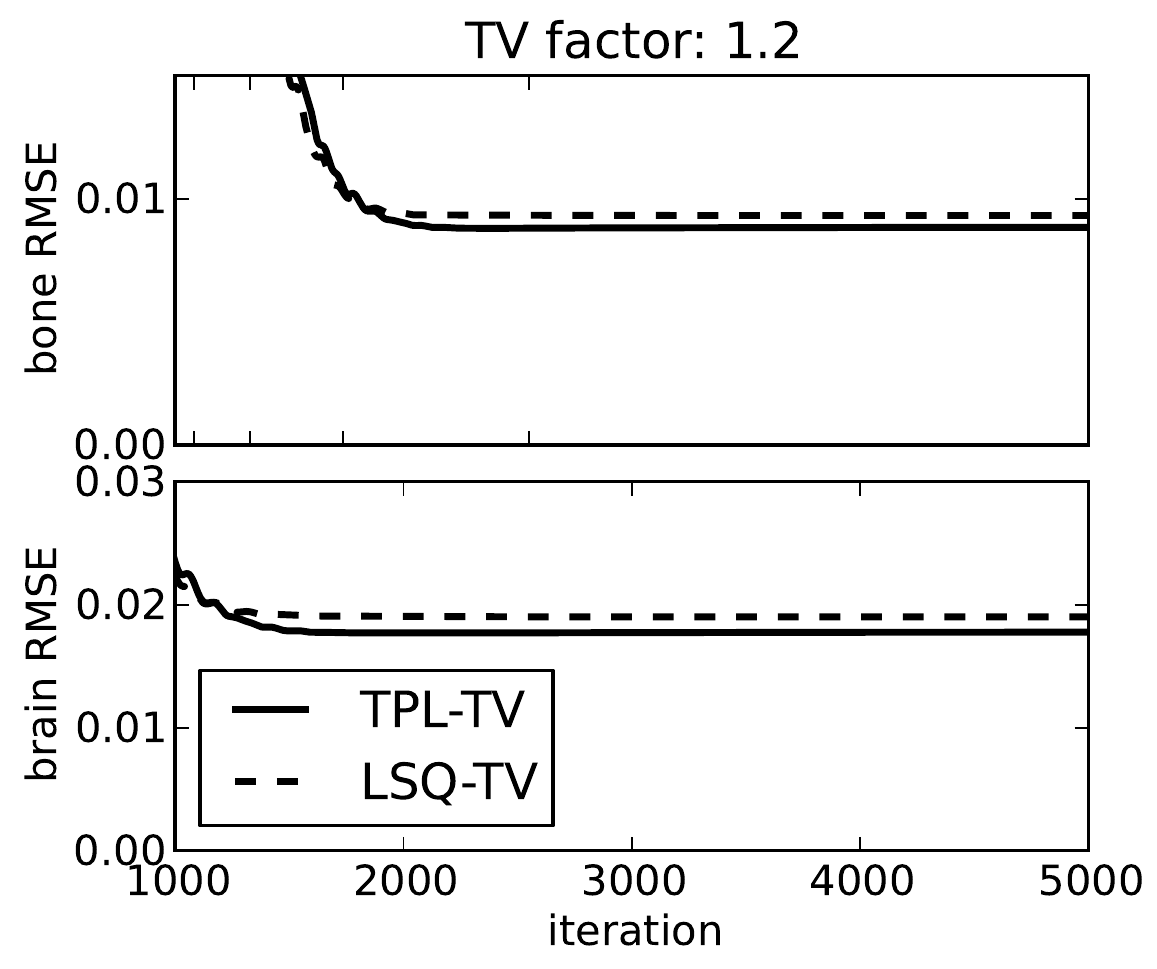}
}
\end{minipage}
\caption{Convergence of the material map estimates to the phantom material maps for
LSQ-TV and TPL-TV and for noisy data with two different settings of the TV constraints.
The TV factor applies to both the bone and brain maps, so that a TV factors of 1.1 and 1.2
correspond to the center and bottom, left images of Figs.~\ref{fig:noisyTPLbone}-\ref{fig:noisyLSQbrain}.
\label{fig:noisyIconvergence}}
\end{figure}

It is instructive to examine the convergence metrics in Fig. \ref{fig:noisyConvergence}
and image convergence in Fig. \ref{fig:noisyIconvergence} for this noisy simulation.
The presentation parallels the noiseless results in Figs. \ref{fig:convergence}
and \ref{fig:iconvergence}, respectively. The differences are that results are shown
for a single $\lambda$ and the image RMSE is shown for two different TV-constraint settings
in the present noisy simulations.  The cPD and TV plots all indicate convergence to
the solution for TPL-TV and LSQ-TV. Note, however, the value of the data discrepancy
objective function settles on a positive value as expected for inconsistent data.
The data discrepancy, however, does not provide a check on convergence. It is true
that if the data discrepancy changes with iteration we do not have convergence, but
the inverse is not necessarily true. It is also reassuring to observe that the convergence
rates for the set values of $\lambda$ are similar between the noiseless and noisy
results. This similarity is also not affected by the fact that the TV constraints are set
to different values in each of these simulations.

The RMSE comparison of the recovered material maps with the true phantom maps shown
in Fig. \ref{fig:noisyIconvergence} indicate an average error less than 1\% for the bone
map and just under 2\% for the brain map (100 $\times$ the RMSE values can be interpreted as a percent
error because the material maps have a value of 0 or 1). The main purpose of showing
these plots is to see quantitatively the difference between the TPL and LSQ data discrepancy
terms. We would expect to see lower values of image RMSE for TPL-TV, because the simulated
noise is generated by a transmission Poisson model. Indeed the image RMSE is lower
for TPL-TV and the gap between TPL-TV and LSQ-TV is larger for looser TV constraints.
We do point out that image RMSE may not translate into better image quality, because
image quality depends on the imaging task for which the images are used. Task-based
assessment would take into account features of the observed signal, noise texture, and possibly
background texture and observer perception \cite{Barrett:FIS}.

One of the benefits of using the TV constraints instead of TV penalties is that the material
maps reconstructed using the TPL and LSQ data discrepancy terms can be compared meaningfully.
The TV constraint parameters will result in material maps with exactly the chosen
TVs, while to achieve the same with the penalization approach the penalty parameters
must be searched to achieve equivalent TVs. Also generating simulation results becomes
more efficient, because we can directly make use of the known phantom TV values.

\subsection{Chest phantom studies with a mono-energetic image TV constraint}
\label{sec:chest}

\begin{figure}[!h]
\begin{minipage}[b]{\linewidth}
\centering
\centerline{
\includegraphics[width=0.8\linewidth]{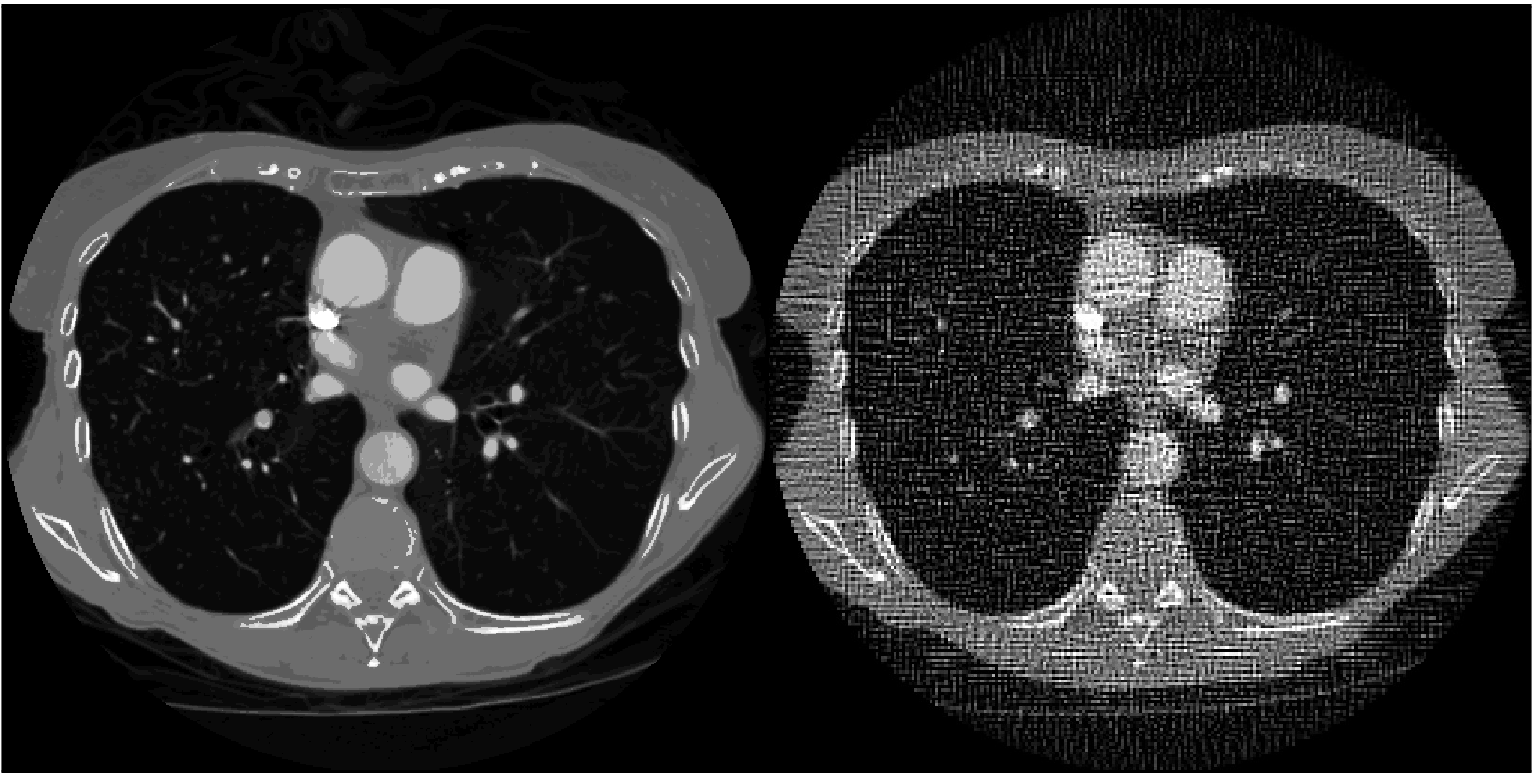}}
\end{minipage}
\caption{(Left) Chest phantom displayed at 70 KeV in a gray scale window of [0, 0.5] cm$^{-1}$.
(Right) Reconstruction by use of unregularized TPL. The estimated
material maps are combined to form the shown monochromatic image estimate at 70 KeV
(gray scale is also [0, 1.0] cm$^{-1}$). For reference the TV values of the phantom
and unconstrained reconstructed image are  2,587 and 7,686, respectively.
\label{fig:chest}}
\end{figure}

For the final set of results we employ an anthropomorphic chest phantom created
from segmentation of an actual CT chest image. Different tissue types and densities
are labeled in the image totaling 24 material/density combinations, including
various soft tissues, calcified/bony regions, and Gadolinium contrast agent. 
To demonstrate the one-step algorithm on this more realistic phantom model,
we select the TPL-monoTV optimization problem in Eq. (\ref{TPL-monoTV}) for the
material map reconstruction. The material basis is selected to be water, bone, and
Gadolinium contrast agent.
Using TPL-monoTV is simpler than TPL-TV in that only the energy for the mono-energetic image and
a single TV constraint parameter is needed 
instead of three parameters -- the TV for each of the material maps. There are potential
advantages to constraining the TV of the material maps individually, but the purpose
here is to demonstrate use of the one-step algorithm and accordingly we select the simpler
optimization problem.  

\begin{figure}[!h]
\begin{minipage}[b]{\linewidth}
\centering
\centerline{
\includegraphics[width=0.55\linewidth]{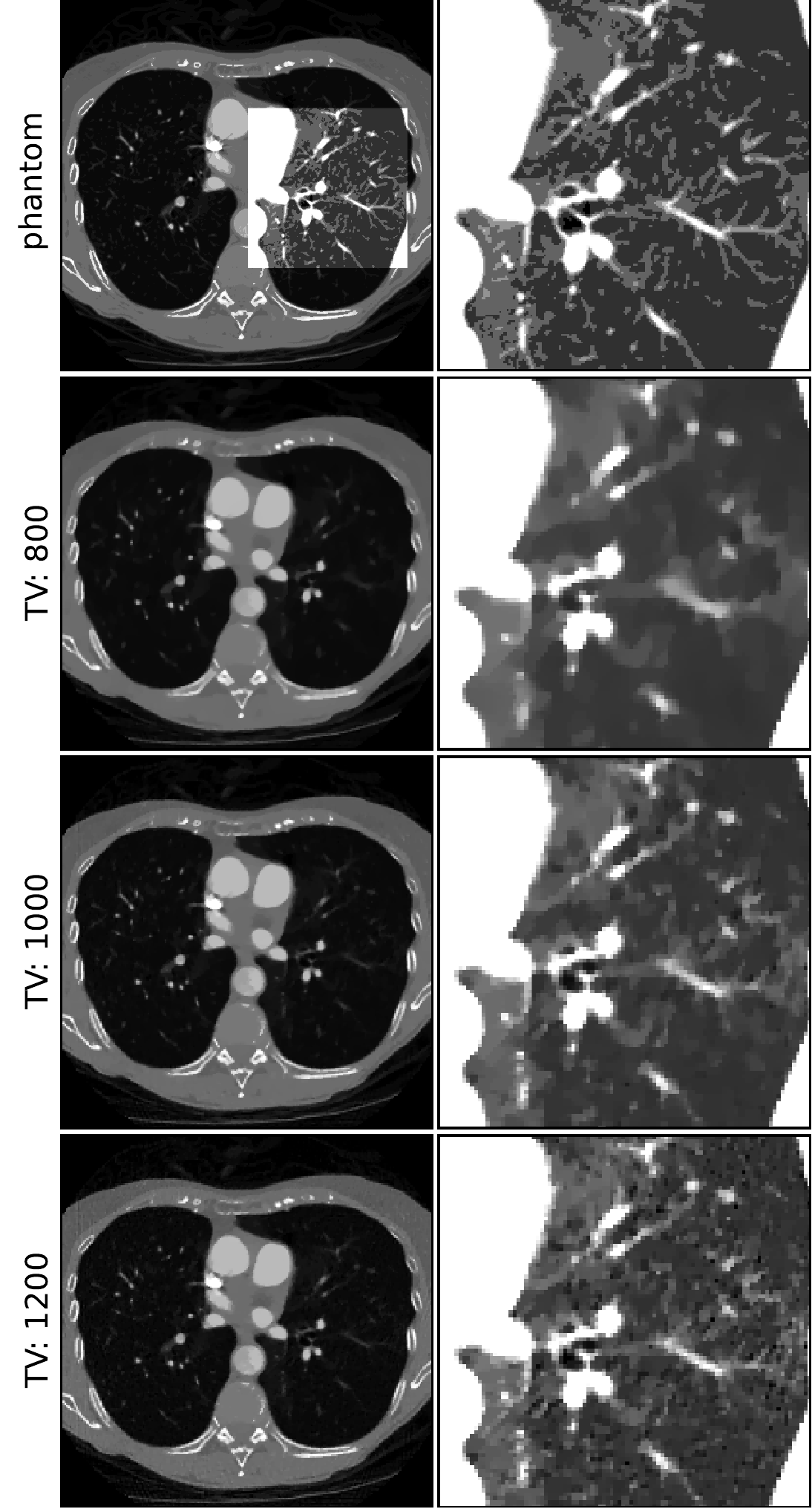}}
\end{minipage}
\caption{Estimated monochromatic images by use of TPL-monoTV. The left
column shows the complete image in a gray scale window of [0, 0.5] cm$^{-1}$.
The right column magnifies a region of interest (ROI)
in the right lung,
and the gray scale is narrowed to [0, 0.1] cm$^{-1}$ in order to see the soft tissue detail.
The top set of images correspond to the phantom.
The location of the ROI is indicated in the left phantom 
image inset by use of the narrow [0, 0.1] cm$^{-1}$ gray scale.
The second, third, and fourth rows
correspond to images obtained by different TV constraints of the monoenergetic image at 70 KeV.
\label{fig:chestResults}}
\end{figure}

For the chest phantom simulations, the scanning configuration is again 2D fan-beam CT with
a source to iso-center distance of 80 cm and source to detector distance of 160 cm. The physical
size of the phantom pixel array is $29 \times 29$ cm$^2$. The number of projection views is
128 over a 2$\pi$ scan.
Five X-ray energy windows are simulated 
in the energy ranges [20 , 50], [50, 60], [60,80], [80,100], and [100, 120] keV.
The lowest energy window is selected wider than the other four to avoid photon starvation.
Noise is added in the same way as the previous simulation. 
The transmitted counts data follow a Poisson model with a total of $4 \times 10^6$ photons
per detector pixel. The monoenergetic image at 70 keV along with unregularized image reconstruction
by TPL are shown in Fig. \ref{fig:chest}. The TPL mono-energetic image reconstruction demonstrates
the impact of the simulated noise on the reconstructed image.

In Fig. \ref{fig:chestResults} we show the resulting monoenergetic images from TPL-monoTV at three
values of the TV constraint. The reconstructed images are shown globally in a wide gray scale and
in an ROI focused on the right lung in a narrow gray scale window. The values of the TV constraint
are selected based on visualization of the fine structures in the lung. For viewing these
features, relatively low values of TV are selected. We note that in the global images the same TV
values show the high-contrast structures with few artifacts. We point out that the one-step
algorithm yields three basis material maps (not shown) and the mono-energetic images are formed by use
of Eq. (\ref{monoimage}).

The selected optimization problems and simulation parameters are chosen to demonstrate possible
applications of the one-step algorithm for spectral CT.
Comparison of the TPL and LSQ data discrepancy
in Figs. \ref{fig:noisyTPLbrain} and \ref{fig:noisyLSQbrain} does show fewer artifacts for
TPL-TV, where the simulated noise model matches the TPL likelihood.
In practice, we may not see the same relative performance on real data --
the simulations ignore
some important physical factors of spectral CT, 
and image quality evaluation depends on the task
for which the images are used.

\section{Conclusion}

We have developed a constrained one-step algorithm for inverting spectral CT transmission data
directly to basis material maps. The algorithm addresses the associated non-convex
data discrepancy terms by employing a local convex quadratic upper bound to derive the descent
step. While we have derived the algorithm for TPL and LSQ data discrepancy terms, the same
strategy can be applied to derive the one-step algorithm for other data fidelities.
The one-step algorithm derives from the convex-concave optimization algorithm, MOCCA,
which we have developed for addressing an intermediate problem arising from
use of the local convex quadratic approximation.
The simulations demonstrate the one-step algorithm for TV-constrained data discrepancy
minimization, where the TV constraints can be applied to the individual basis maps or to
an estimated monochromatic X-ray attenuation map.

Future work will investigate robustness of the one-step algorithm to data inconsistency
due to spectral miscalibration error, X-ray scatter, and various physical processes
involved in photon-counting detection.  The one-step algorithm's ability to incorporate
basis map constraints in the inversion process should provide a means to control artifacts
due to such inconsistencies. We are also pursuing a generalization to the present algorithm
to allow for auto-calibration of the spectral response of the CT system.

\section*{Acknowledgements}
This work was supported by NIH Grants R21EB015094, CA158446, CA182264, and EB018102.
The contents of this article are solely the responsibility of the authors and do not necessarily
represent the official views of the National Institutes of Health. 

\appendices
\section{Gradient of $L_\text{TPL}$}
\label{sec:derivatives}

We derive the gradient in Eq. (\ref{TPLgrad}), motivating the definition of the linear transform
$A$.
Recall Eqs. (\ref{meanModel}) and (\ref{zdef}):
\begin{linenomath}
\begin{align}
\notag
Z_{\ell i, m k}& =  \mu_{m i} X_{\ell k} \\
\notag
\hat{c}_{w \ell}(f_{k m}) & = N_{w \ell} \sum_i s_{w \ell i}
\exp \left[ -\sum_{m k} \mu_{m i} X_{\ell k} f_{k m} \right] \\
\notag
& = N_{w \ell} \sum_i s_{w \ell i}
\exp \left[ -\sum_{m k} Z_{\ell i, m k} f_{k m} \right] \\
\notag
& = N_{w \ell} \sum_i s_{w \ell i}
\exp \left[ - (Zf)_{\ell i}  \right].
\end{align}
\end{linenomath}
The gradient of $L_\text{TPL}$ is
\begin{linenomath}
\begin{align}
\notag
\nabla_f L_\text{TPL}(f) &:= \frac{\partial}{\partial f_{km}} \sum_{w \ell} \left[ \hat{c}_{w \ell} (f) - c_{w \ell} -
c_{w \ell} \log \left( \hat{c}_{w \ell} (f)/ c_{w \ell} \right) \right] \\
\notag
&=  \frac{\partial}{\partial f_{km}} \sum_{w \ell} \left[\hat{c}_{w \ell}(f_{km}) - 
c_{w \ell} \log \hat{c}_{wl}(f_{km}) \right] \\
\notag
&= \sum_{w \ell} \left [N_{w \ell} \sum_i s_{w \ell i} \cdot(-Z_{\ell i, m k}) \exp[-(Zf)_{\ell i}]
-c_{w \ell} \frac{\sum_i s_{w \ell i} \cdot(-Z_{\ell i, m k}) \exp[-(Zf)_{\ell i}]}
{\sum_i s_{w \ell i} \exp[-(Zf)_{\ell i}]} \right] \\
\notag
&=\sum_{\ell i} Z_{\ell i, m k} \sum_w
\left[ c_{w \ell} \frac{ s_{w \ell i} \exp[-(Zf)_{\ell i}]}
{\sum_i s_{w \ell i} \exp[-(Zf)_{\ell i}]} -
N_{w \ell}  s_{w \ell i} \exp[-(Zf)_{\ell i}] \right]\\
\notag
&=\sum_{\ell i} Z_{\ell i, m k} \sum_w
(c_{w \ell}-\hat{c}_{w \ell} (f)) \frac{ s_{w \ell i} \exp[-(Zf)_{\ell i}]}
{\sum_i s_{w \ell i} \exp[-(Zf)_{\ell i}]} .
\end{align}
\end{linenomath}
Continuing the algebraic manipulation we insert $\mathbf{I}_{\ell \ell^\prime}$:
\begin{linenomath}
\begin{align}
\notag
\nabla_f L_\text{TPL}(f) & =\sum_{\ell i} Z_{\ell i, m k} \sum_{w \ell^\prime}
\mathbf{I}_{\ell \ell^\prime}
\frac{ s_{w \ell^\prime i} \exp[-(Zf)_{\ell^\prime i}]}
{\sum_i s_{w \ell^\prime i} \exp[-(Zf)_{\ell^\prime i}]}(c_{w \ell^\prime}-\hat{c}_{w \ell^\prime} (f)) \\
\notag
&= \sum_{\ell i} Z_{\ell i, m k} \sum_{w \ell^\prime} A_{w \ell^\prime, \ell i}(f) r_{w \ell^\prime}(f)
= Z^\top A(f)^\top r(f).
\end{align}
\end{linenomath}
The other necessary gradient and Hessian computations follow from similar manipulations.

\section{Positive semidefiniteness of $\nabla^2_+ L(f)$ and $\nabla^2_- L(f)$}
\label{sec:posdef}

Recall Eq. (\ref{adef}),
\begin{linenomath}
\begin{equation}
\notag
A_{w \ell, \ell^\prime i}(f) =  \frac{s_{w \ell i} \exp \left[- (Zf)_{\ell i} \right]}
{\sum_{i^\prime} s_{w \ell i^\prime} \exp \left[- (Zf)_{\ell i^\prime} \right]}
 \mathbf{I}_{\ell^\prime \ell}.
\end{equation}
\end{linenomath}
For ease of presentation, we collapse the double indices of $A$
\begin{linenomath}
\begin{equation}
\notag
A_{s,t}(f) = A_{w \ell, \ell^\prime i}(f) \text{ where } 
s = w \cdot N_\ell + \ell,\;t=i \cdot N_\ell + \ell^\prime.
\end{equation}
\end{linenomath}
We show that
\begin{linenomath}
\begin{equation}
\label{ineq1}
\diag \left( A(f)^\top b \right) -A(f)^\top \diag (b) A(f) \geq 0
\text{ when } b_s \geq 0 \; \forall \, s .
\end{equation}
\end{linenomath}
This inequality can be used to prove that the Hessians in Eqs. (\ref{TPLposH}),
(\ref{TPLnegH}), (\ref{LSQposH}), and (\ref{LSQnegH}) are positive semidefinite
by setting $b$ equal to $r_-(f)$, $r_+(f)$, $r_-^\text{(log)}(f)$, and $r_+^\text{(log)}(f)$,
respectively.

To prove Eq. (\ref{ineq1}), we expand $b$ in unit vectors
\begin{linenomath}
\begin{equation}
\notag
b = \sum_s b_s \hat{e}_s \text{ where } \hat{e}_{s,s^\prime} =
\begin{cases}
1 & s^\prime = s\\
0 & s^\prime \neq s
\end{cases}.
\end{equation}
\end{linenomath}
and show for any vector $u$ that
\begin{linenomath}
\begin{equation}
\label{ineq2}
u^\top \left[ \diag \left( A(f)^\top \hat{e}_s \right)
-A(f)^\top \diag( \hat{e}_s) A(f) \right] u \geq 0.
\end{equation}
\end{linenomath}
Fixing $s$, we define the vector $v$ with components
$v_t = A(f)_{s,t}$ or using the unit vector $\hat{e}_s$
\begin{linenomath}
\begin{equation}
\notag
v = A^\top(f) \hat{e}_s.
\end{equation}
\end{linenomath}
From the definition of $A$, we note that $v_t \geq 0$ for all $t\in \{1,\dots,N_t\}$ and $\sum_t v_t = 1$.
By the definition of $v$, the left-hand side of Eq. (\ref{ineq2}), lhs,
is 
\begin{linenomath}
\begin{align}
\notag
\text{lhs} &= u^\top \left[\diag(v) - vv^\top \right] u \\
\notag
           &= \sum_t v_t u_t^2 -\left( \sum_t v_t u_t \right)^2\\
\notag
           &= \sum_t v_t u_t^2 -\left( \sum_t \sqrt{v_t} \cdot \sqrt{v_t u_t^2} \right)^2,
\end{align}
\end{linenomath}
and by the Cauchy-Schwartz inequality,
\begin{linenomath}
\begin{align}
\notag
\text{lhs} &\geq
\sum_t v_t u_t^2 -\left( \sum_t v_t \right) \cdot
\left( \sum_t v_t u_t^2 \right) \\
\notag
&=
\sum_t v_t u_t^2 -(1) \cdot
\left( \sum_t v_t u_t^2 \right) \\
\notag
& = 0.
\end{align}
\end{linenomath}
This proves the inequality in Eq. (\ref{ineq2}).

Using Eq. (\ref{ineq2}), we prove the inequality in Eq. (\ref{ineq1})
\begin{linenomath}
\begin{equation}
\diag \left( A(f)^\top b \right) -A(f)^\top \diag (b) A(f) 
= \sum_s b_s \cdot \left[ \diag \left( A(f)^\top \hat{e}_s \right)
-A(f)^\top \diag( \hat{e}_s) A(f) \right] \geq 0,
\end{equation}
\end{linenomath}
where the inequality is shown by noting that $b_s \geq 0$, by assumption,
and the sum is thus a linear combination of positive definite matrices with
non-negative coefficients.

\section{Derivation of one-step algorithm for TPL-TV and LSQ-TV with $\mu$-preconditioning}
\label{sec:onestepalgorithm}

\paragraph*{TV-constrained optimization}
To derive the one-step algorithm used in the Results section, we write
down the intermediate convex optimization problem that involves the 
first block of the local
quadratic upper bound to $D_\text{TPL}$ or $D_\text{LSQ}$
\begin{linenomath}
\begin{equation}
\label{TPL-TValg_opt}
f^\star = \argmin_f \left\{ \frac{1}{2} (K_1 f)^\top D_1 K_1 f - (K_1 f-z_0)^\top (b_1+E_1 z_0)
\right\} \text{ such that } \| f_m \|_\text{TV} \le \gamma_m \; \forall \, m .
\end{equation}
\end{linenomath}
That only the first block of the full quadratic expression is explained in Sec. \ref{sec:onestepMOCCA}
and the form of $D_1$, $E_1$ and $b_1$ given in Sec. \ref{sec:cpgen} determines
whether we are addressing TPL-TV or LSQ-TV .
The data discrepancy term of this optimization problem is the same
as the objective function of Eq. (\ref{FQconvex}),
but it differs from Eq. (\ref{FQconvex}) in that we have added the
convex constraints on the material map TV values. We write Eq. (\ref{TPL-TValg_opt})
using indicator functions (see Eq. (\ref{indicatordef}))
to code the TV constraints and we introduce the $\mu$-preconditioning
transformation described in Sec. \ref{sec:preconditioning}
\begin{linenomath}
\begin{multline}
\label{TPL-TValg_opt2}
f^\star = \argmin_{f^\prime} \left\{ \frac{1}{2} (K^\prime_1 f^\prime)^\top D_1 K^\prime_1 f -
(K^\prime_1 f^\prime -z_0)^\top (b_1 +E_1 z_0) +\right. \\
\left. \sum_m \delta \left( 
\left\| \left( P^{-1} f^\prime  \right)_m
\right\|_\text{TV} \le \gamma_m \right) \right\},
\end{multline}
\end{linenomath}
where $f^\prime = P f$ are the transformed ($\mu$-preconditioned)
material maps from Sec. \ref{sec:preconditioning}.
Note that the TV constraints apply to the untransformed material maps $f = P^{-1} f^\prime$.

\paragraph*{Writing constrained TV optimization in the general form $F(Kx)+G(x)$}
To derive the CP primal-dual algorithm, we write Eq. (\ref{TPL-TValg_opt2}) in the
form of Eq. (\ref{cpgen}). We note that all the terms involve a linear transform of $f$, and
accordingly we make the following assignments
\begin{linenomath}
\begin{align}
\notag
F_\text{convex}(z_0;z) &= F_1(z_0; z_\text{sino}) +F_2(z_\text{grad})\\
\notag
F_1(z_\text{sino}) &= \frac{1}{2} z^\top_\text{sino} D_1 z_\text{sino} -(z_\text{sino}-z_0)^\top ( b_1+E_1 z_0)\\
\notag
F_2(z_\text{grad}) &= \sum_m \delta \left(
\left\| \left( | z_{\text{grad},m} |  \right) \right\|_1 \le \gamma_m \right) \\
\notag
G(f^\prime) &= 0,
\end{align}
\end{linenomath}
where
\begin{linenomath}
\begin{equation}
\notag
z = \left( 
 \begin{array}{c}
 z_\text{sino} \\
 z_\text{grad}
 \end{array} \right) =
 \left( 
 \begin{array}{c}
 K^\prime_1 \\
 \nabla P^{-1}
 \end{array} \right) f^\prime.
\end{equation}
\end{linenomath}
Note that we use the short-hand that the gradient operator, $\nabla$, applies to each of the material maps 
in the composite material map vector, $P^{-1} f^\prime$.
The Legendre transform in Eq. (\ref{legendre}) provides the necessary dual functions
$F^*_1$, $F^*_2$, and $G^*$.
By direct computation
\begin{linenomath}
\begin{equation}
\label{F1star}
F^*_1(y_\text{sino}) = \frac{1}{2} D_1^{-1} \| y_\text{sino} +b_1 + E_1 z_0 \|^2_2 +z_0^T E_1 z_0.
\end{equation}
\end{linenomath}
From Sec. 3.1 of Ref. \cite{Sidky2012}
\begin{linenomath}
\begin{equation}
\label{Gstar}
G^*(f) =\delta( f=0).
\end{equation}
\end{linenomath}

\begin{figure}[!h]
\begin{minipage}[b]{\linewidth}
\centering
\centerline{
\includegraphics[width=0.3\linewidth]{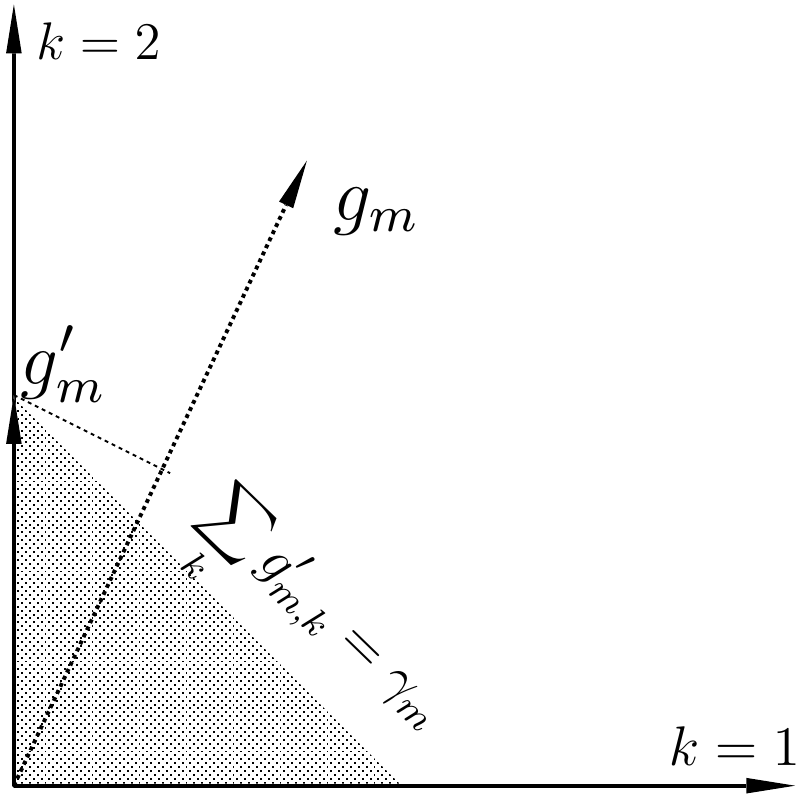}}
\end{minipage}
\caption{Schematic illustrating the solution of
$ \max_{g_m^\prime} \left\{ g_m^\top g_m^\prime - \delta \left(
\left\|  g^\prime_{m}  \right\|_1 \le \gamma_m \right) \right\}$.
The input vector $g_m$ and the maximizing vector $g^\prime_m$ are indicated on a 2D
schematic, but the argument applies for the full $N_k$-D space of $g_m$.
Because $g^\prime_m$ is a vector of magnitudes, each component is non-negative
$g^\prime_{m,k} \geq 0$.  The indicator function confines $g^\prime_m$ below
the line (hyper-plane), $\sum_k g^\prime_{m,k} = \gamma_m$. The combination
of these constraints confines $g^\prime_m$ to the schematic, shaded triangle.
The maximizer $g^\prime_m$ is the vector that maximizes the dot product,
$g_m^\top g_m^\prime$ (or equivalently the projection of $g^\prime_m$ onto
$g_m$ as indicated by the dashed line from the head of $g^\prime_m$ to the
arrow indicating $g_m$). Maximization of this dot product is achieved by
choosing $g^\prime_m = \gamma_m \hat{e}_{k-\text{max}}$
 such that it is aligned along the unit vector corresponding
to the largest component of $g_m$. The largest component of $g_m$ is also known as
the ``infinity-norm'', $\| g_m \|_\infty$. Thus we have
$(\gamma_m \hat{e}_{k-\text{max}})^\top g_m= \gamma_m \|g_m\|_\infty$.
\label{fig:dtvdual}}
\end{figure}

\paragraph*{Convex conjugate of $F_2$}
We sketch the derivation of $F^*_2(y_\text{grad})$, and for this derivation we
drop the ``grad'' subscript.
\begin{linenomath}
\begin{align}
\notag
F^*_2(y) &= \max_{y^\prime} \left\{ y^\top y^\prime - \sum_m \delta \left(
\left\| \left( | y^\prime_{m} |  \right) \right\|_1 \le \gamma_m \right) \right\} \\
\notag
&= \max_{g^\prime} \left\{ g^\top g^\prime - \sum_m \delta \left(
\left\|  g^\prime_{m}  \right\|_1 \le \gamma_m \right) \right\} \\
\notag
&\text{where } g = |y| \text{ and } g^\prime = |y^\prime|,\\
\notag
&= \max_{g^\prime} \sum_m \left\{ g_m^\top g_m^\prime - \delta \left(
\left\|  g^\prime_{m}  \right\|_1 \le \gamma_m \right) \right\},\\
\notag
&= \sum_m \max_{g_m^\prime} \left\{ g_m^\top g_m^\prime - \delta \left(
\left\|  g^\prime_{m}  \right\|_1 \le \gamma_m \right) \right\}
\end{align}
\end{linenomath}
The Legendre transform maximization over the variable $y^\prime$, dual to the material map
gradients, is reduced to a maximization over the spatial magnitude $g^\prime = |y^\prime|$
because the indicator function is independent of the spatial direction of $y^\prime$
and the term $y^\top y^\prime$ is maximized when the spatial direction of $y^\prime$ line up
with $y$; hence the term $y^\top y^\prime$ is replaced by $g^\top g^\prime$, which we explicitly
write as a sum over the material index $m$. The maximization and summation order can be switched,
because each of the terms in the summation are independent of each other.
Evaluation of the maximization over $g^\prime_m$ can be seen in the diagram shown in Fig. \ref{fig:dtvdual}.
Accordingly we find
\begin{linenomath}
\begin{equation}
\label{F2star}
F^*_2(y_\text{grad}) = \sum_m \gamma_m \| (|y_\text{grad}|) \|_\infty.
\end{equation}
\end{linenomath}

\paragraph*{Dual maximization of Eq. (\ref{TPL-TValg_opt2})}
Using Eqs. (\ref{cpgen_dual}), (\ref{F1star}), (\ref{Gstar}), and (\ref{F2star}), we
obtain the maximization dual to Eq. (\ref{TPL-TValg_opt})
\begin{linenomath}
\begin{multline}
\label{TPL-TValg_opt2_dual}
y^\star =\argmax_y \left\{ -\frac{1}{2} D^{-1}_1 \| y_\text{sino}+b_1 +E_1 z_0 \|_2^2 - z^\top_0 E_1 z_0
-\sum_m \gamma_m \| (|y_\text{grad}|) \|_\infty \right\} \\
\text{ such that }
\left( \begin{array}{c} K^\prime_1 \\ \nabla P^{-1} \end{array} \right)^\top y=0.
\end{multline}
\end{linenomath}
The objective functions of the primal and dual problems, in Eqs. (\ref{TPL-TValg_opt})
and (\ref{TPL-TValg_opt2_dual}) respectively, are needed to generate the conditional
primal-dual gap plots in Fig. \ref{fig:convergence}.

\paragraph*{The material map TV proximity step}
In order to derive the TPL-TV and LSQ-TV one step algorithms, we need to derive
the proximity minimization in Eq. (\ref{moccadualstep})
\begin{linenomath}
\begin{equation}
\notag
y^{(n+1)}  =  \argmin_{y^\prime} \left\{F_\text{convex}^*\left(z^{(n+1)}_0; y^\prime\right)
+ \frac{1}{2} \left\| \Sigma_\text{grad}^{-1/2}
\left( y^{(n)}+(\Sigma_\text{sino} K^\prime_1 +
\Sigma_\text{grad} \nabla P^{-1}) \bar{x}^{(n)} - y^\prime
 \right) \right\|^2_2
\right\}.
\end{equation}
\end{linenomath}
The proximity problem splits into ``sino'' and ``grad'' sub-problems and the ``sino'' sub-problem
results in Eq. (\ref{sct-moccadualstep}). We solve here the ``grad'' proximity optimization
to obtain the pseudo-code for TPL-TV and LSQ-TV
\begin{linenomath}
\begin{equation}
\notag
y_\text{grad}^{(n+1)}  =  \argmin_{y_\text{grad}^\prime}
\left\{F_2^*\left(y_\text{grad}^\prime\right)
+ \frac{1}{2 } \left\| \Sigma_\text{grad}^{-1/2} \left(   y^+_\text{grad} -
y_\text{grad}^\prime \right) \right\|^2_2 \right\} 
\text{ where } y^+_\text{grad} = y_\text{grad}^{(n)}+\Sigma_\text{grad} \nabla P^{-1} \bar{x}^{(n)}.
\end{equation}
\end{linenomath}
Dropping the ``grad'' subscript on $y^\prime$, we employ the Moreau identity which relates the proximity
optimizations between a function and its dual
\begin{linenomath}
\begin{multline}
\notag
\argmin_{y^\prime} \left\{F_2^*\left(y^\prime\right) + \frac{1}{2}
\left\|\Sigma_\text{grad}^{-1/2} \left(y^+_\text{grad} - y^\prime \right) \right\|^2_2 \right\}
  = \\
 y^+_\text{grad}
 - \Sigma_\text{grad}^{1/2}  \argmin_{y^\prime}
\left\{F_2\left(y^\prime \Sigma_\text{grad}^{-1/2} \right) +
\frac{1}{2} \left\| 
 y^+_\text{grad} \Sigma_\text{grad}^{-1/2}  - y^\prime  \right\|^2_2 \right\} .
\end{multline}
\end{linenomath}
The dual ``grad'' update separates into the individual material map $m$ components
\begin{linenomath}
\begin{equation}
\notag
y_{\text{grad},m}^{(n+1)}  = y^+_{\text{grad},m}
 - \Sigma_{\text{grad},m}^{1/2}  \argmin_{y^\prime_m}
\left\{
\delta \left(
\left\| \left( \left|y^\prime_m\Sigma_{\text{grad},m}^{-1/2} \right|\right) \right\|_1 \le \gamma_m \right) +
\frac{1}{2} \left\| 
 y^+_{\text{grad},m}\Sigma_{\text{grad},m}^{-1/2}  - y^\prime_m  \right\|^2_2 \right\}.
\end{equation}
\end{linenomath}
To simplify the proximity minimization we
set
\begin{linenomath}
\begin{align}
\notag
g^+_m &= \left|y^+_{\text{grad},m}\Sigma_{\text{grad},m}^{-1/2} \right|, \;\;
\hat{g}^+_m = \left(y^+_{\text{grad},m}\Sigma_{\text{grad},m}^{-1/2} \right)/g^+_m, \;\;
g^\prime_m = \left|y^\prime_m\right|,\\
\notag
y_{\text{grad},m}^{(n+1)}  &= y^+_{\text{grad},m}
 - \Sigma_{\text{grad},m}^{1/2} \, \hat{g}^+_m\argmin_{g^\prime_m}
\left\{
\delta \left(
\left\|  g^\prime_m\Sigma_{\text{grad},m}^{-1/2} \right\|_1 \le \gamma_m \right) +
\frac{1}{2} \left\| 
 g^+_m  - g^\prime_m  \right\|^2_2 \right\}.
\end{align}
\end{linenomath}
The proximity minimization is a projection of $g^+_m$ onto a weighted $\ell_1$-ball.
\begin{linenomath}
\begin{align}
\notag
y_{\text{grad},m}^{(n+1)} &= y^+_{\text{grad},m}
 - w \, \hat{g}^+_m 
\text{Proj} \left( g^+_m; \left\{ g,\|g/w\| \le \gamma_m \right\} \right) \\
\notag
& \text{where } w = \Sigma_{\text{grad},m}^{1/2}.
\end{align}
\end{linenomath}

If $g$ is inside the weighted $\ell_1$-ball, i.e. $ \| g/w \|_1 \leq \gamma$,
the function $\text{Proj} \left( g; \left\{ g,\|g/w\| \le \gamma \right\} \right)$ returns $g$.
If $g$ is outside the weighted $\ell_1$-ball, i.e. $ \| g/w \|_1 > \gamma$,
there exists an $\alpha_0$ such that
\begin{linenomath}
\begin{align}
\notag
\text{Proj} \left( g; \left\{ g,\|g/w\| \le \gamma_m \right\} \right) &=
\argmin_{g^\prime} 
\left\{
\delta \left(
\left\|  g^\prime/w \right\|_1 \le \gamma \right) +
\frac{1}{2} \left\| 
 g  - g^\prime  \right\|^2_2 \right\} \\
\notag
& = \argmin_{g^\prime}
\left\{ \alpha_0 \left\|  g^\prime/w \right\|_1 
 +
\frac{1}{2} \left\| 
 g  - g^\prime  \right\|^2_2 \right\}\\
\notag
& = \max \left( g - \alpha_0 w  , 0 \right).
\end{align}
\end{linenomath}
The parameter $\alpha_0$ is defined implicitly
\begin{linenomath}
\begin{equation}
\notag
\| \max \left( g - \alpha_0 w  , 0 \right) \|_1 = \gamma,
\end{equation}
\end{linenomath}
and it can be determined by any standard root finding technique applied
to
\begin{linenomath}
\begin{equation}
\notag
f(\alpha) =0 \text{ where } f(\alpha) = \| \max \left( g - \alpha w  , 0 \right) \|_1 - \gamma,
\end{equation}
\end{linenomath}
where the search interval is $\alpha \in [0, \|g/w\|_\infty]$.

\paragraph*{The pseudocode for TPL-TV and LSQ-TV}
Having derived the TV constraint proximity step, 
we are in a position to write the complete pseudocode for the one-step algorithm  including the
TV constraints. We do employ the $\mu$-preconditioning that orthogonalizes the linear attenuation
coefficients, but we drop the prime notation on $f$ and $K$.
\begin{linenomath}
\begin{align}
\notag
f_0 &= \bar{f}^{(n)} \\
\notag
\Sigma^{(n)} &= \left(
\begin{array}{c} |K_1(f_0)|\\ |\nabla P^{-1}| \end{array}
\right) \mathbf{1}/\lambda ; \; \;
T^{(n)} = \lambda  \left(
\begin{array}{c} |K_1(f_0)|\\ |\nabla P^{-1}| \end{array}
\right)^\top \mathbf{1} \\
\notag
w_m &= \left(\Sigma_{\text{grad},m}^{(n)}\right)^{1/2}  \; \forall \, m \\
\notag
z^{(n+1)}_0 &=
\left( \Sigma_\text{sino}^{(n)} \right)^{-1}
\left(y_\text{sino}^{(n-1)} - y_\text{sino}^{(n)} +\Sigma_\text{sino}^{(n)}  K_1(f_0) \bar{f}^{(n-1)} \right) \\
\notag
y_\text{sino}^{(n+1)} & = \left(D_1(f_0) + \Sigma_\text{sino}^{(n)}\right)^{-1}  \left[ D_1(f_0)
\left( y_\text{sino}^{(n)} + \Sigma_\text{sino}^{(n)}  K_1(f_0) \bar{f}^{(n)} \right) -
 \Sigma_\text{sino}^{(n)} \left( b_1(f_0) + E_1(f_0) z^{(n+1)}_0 \right) \right] \\
\notag
y^+_{\text{grad},m} &= y_{\text{grad},m}^{(n)}+\Sigma_{\text{grad},m}^{(n)} \nabla (P^{-1} \bar{f}^{(n)})_m
\; \forall \, m\\
\notag
g^+_m &= \left|y^+_{\text{grad},m}/w_m\right|; \;\;
\hat{g}^+_m = \left(y^+_{\text{grad},m}/w_m\right)/g^+_m \; \forall \, m  \\
\notag
y_{\text{grad},m}^{(n+1)} &= y^+_{\text{grad},m}
 - w_m \, \hat{g}^+_m 
\text{Proj} \left( g^+_m; \left\{ g,\|g/w_m\| \le \gamma_m \right\} \right) \; \forall \, m \\
\notag
f^{(n+1)} & =  \bar{f}^{(n)} - T^{(n)}  \left(
\begin{array}{c} K_1(f_0)\\ \nabla P^{-1} \end{array}
\right)^\top \left( \begin{array}{c} y_\text{sino}^{(n+1)} \\ y_\text{grad}^{(n+1)} \end{array} \right)\\
\notag
\bar{f}^{(n+1)} &=  2f^{(n+1)} -f^{(n)}
\end{align}
\end{linenomath}
The final material maps after $N$ iterations are obtained by applying the inverse preconditioner
\begin{linenomath}
\begin{equation}
\notag
\text{return } P^{-1} f^{(N)}.
\end{equation}
\end{linenomath}
For all the results presented in the article, all variables are initialized to zero.

\bibliography{one_step_intro,one_step_body}
\bibliographystyle{IEEEtran}

\end{document}